\documentclass[12pt,preprint]{aastex}

\newcommand{\Msol}{\rm{M}_{\sun}}

\newcommand{\sh}{\mbox{Sh\,2-174}}
\newcommand{\masyr}{\mbox{mas\,yr$^{-1}$}}
\newcommand{\kms}{\mbox{km\,s$^{-1}$}}
\newcommand{\cmss}{\mbox{cm\,s$^{-2}$}}

\slugcomment{Accepted for publication in ApJ.}

\shorttitle{A Radio Investigation of \sh}
\shortauthors{Ransom et al.}

\begin{document}

\title{THE EMISSION NEBULA \sh\@: A RADIO INVESTIGATION OF THE
  SURROUNDING REGION}

\author{R.\ R.\ Ransom\altaffilmark{1,2}, R.\ Kothes\altaffilmark{2},
  J.\ Geisbuesch\altaffilmark{2}, W.\ Reich\altaffilmark{3}, and
  T.\ L.\ Landecker\altaffilmark{2}}

\altaffiltext{1}{Department of Physics and Astronomy, Okanagan
  College, 583 Duncan Avenue West, Penticton, B.C., V2A 8E1, Canada}

\altaffiltext{2}{National Research Council Herzberg, Dominion Radio
  Astrophysical Observatory, P.O.\ Box 248, Penticton, BC, V2A 6J9,
  Canada}

\altaffiltext{3}{Max-Planck-Institut f{\"u}r Radioastronomie, Auf dem
  H{\"u}gel 69, D-53121 Bonn, Germany}


\begin{abstract}

\sh\@ is believed to be either a planetary nebula (PN) or ionized,
ambient interstellar medium (ISM).  We present in this paper 1420~MHz
polarization, 1420~MHz total intensity (Stokes-$I$), and neutral
hydrogen (\ion{H}{1}) images of the region around \sh\@.  The radio
images address not only the nature of the object, but also the history
of the relationship between \sh\@ and its surrounding environment.
The \ion{H}{1} images show that \sh\@ sits presently at the center of
a $\sim$$1.2\arcdeg$\,$\times$\,$\sim$$0.4\arcdeg$ cloud (with peak
hydrogen density $n_H = 4 \pm 2$~$\rm{cm}^{-3}$).  The Stokes-$I$
image shows thermal-emission peaks (with electron densities $n_e = 11
\pm 3$~$\rm{cm}^{-3}$) coincident with the R-band optical nebula, as
well as low-surface-brightness emission from an ionized ``halo''
around \sh\@ and from an ionized ``plateau'' extending southeast from
the cloud.  The polarization images reveal Faraday-rotation structures
along the projected trajectory of \sh\@, including a high-contrast
structure with ``arms'' that run precisely along the eastern edge of
the \ion{H}{1} cloud and a wide central region which merges with the
downstream edge of \sh\@.  The high-contrast structure is consistent
with an ionized tail which has both {\it early-epoch} (before \sh\@
entered the cloud) and {\it present-epoch} (after \sh\@ entered the
cloud) components.  Furthermore, our rotation-measure analysis
indicates that the ISM magnetic field is deflected at the leading edge
of \sh\@.  The downstream tail and upstream field deflection point to
a PN-ISM interaction.  Our estimated space velocity for the host white
dwarf (GD~561) demonstrates that \sh\@ entered the cloud
$\sim$27,000~yr ago, and gives a PN-ISM interaction timescale
$\lesssim$$2.0 \times 10^5$~yr.  We estimate an ambient magnetic field
in the cloud of $11 \pm 3$~$\mu\rm{G}$.

\end{abstract}

\keywords{planetary nebulae: individual (\sh\@) --- stars: individual
  (GD~561) --- stars: kinematics and dynamics --- ISM: structure ---
  ISM: clouds --- ISM: magnetic fields --- radio lines: ISM ---
  techniques: interferometric --- techniques: polarimetric}

\section{INTRODUCTION\label{intro}}

\sh\@ (LBN~598)\footnote{The emission nebula was discovered by
  \citet{Sharpless1959} and also noted by \citet{Lynds1965}.} was
first identified as a planetary nebula (PN) by
\citet*{NapiwotzkiS1993} on account of its apparent association with
the hot white dwarf GD~561 (WD\,2342+806).  Shortly thereafter,
\citet*{TweedyN1994} confirmed the physical association between \sh\@
and GD~561, and, based on the significant westward offset of GD~561
from the center of the nebulosity (see Section~\ref{spacevel}), termed
\sh\@ ``the planetary nebula abandoned by its central star.''
\citet*{TweedyK1996} included \sh\@ in their atlas of ancient PNe, and
noted that the displaced ``central'' star and sharp western edge of
the \ion{O}{3} region \citep*[see][]{TweedyN1994} demonstrate two of
the principal signs of an advanced interaction between a PN and the
interstellar medium (ISM).  The general classification of \sh\@ as an
interacting PN has continued to the present day \citep[see,
  e.g.,][]{Ali+2012,Ali+2013}.

\citet*{FrewP2010} offer an opposing view of the nature of \sh\@, and
other low-surface-brightness emission nebulae generally considered to
be old PNe (see, also, \citealt{Kohoutek1983}; \citealt*{AckerS1990};
\citealt{Madsen+2006}; \citealt{Parker+2006}; \citealt{Frew+2010}).
They believe \sh\@ to be a Str{\"o}mgren zone; i.e., ambient ISM
material, ionized and thermally excited by GD~561.  This conclusion is
based on the absence of an enhanced rim of emission at the leading
edge of \sh\@, the relatively narrow widths of the \sh\@ emission
lines, and differences in the kinematics between the ionized gas and
GD~561.  An additional piece of evidence in favor of a
Str{\"o}mgren-zone interpretation is the evolutionary status of
GD~561.  Assuming a standard post-asymptotic-giant-branch (post-AGB)
evolution, GD~561 is $>$$10^6$~yr old \citep[e.g.,][]{Blocker1995b}.
This age estimate is a factor $>$4 larger than the estimated ages of
even the oldest interacting PNe \citep[see][]{Ali+2012}.

Three-dimensional hydrodynamic simulations of the PN-ISM interaction
reveal structures that should help distinguish an interacting PN from
a Str{\"o}mgren zone (\citealt*{WareingZO2007main}).  At the early
stages of the interaction, a bow shock is formed,\footnote{Simulations
  show that the bow shock is actually created during the earlier AGB
  phase (see, also, \citealt{Wareing+2006agbbow};
  \citealt{Wareing+2007mira}; \citealt{Cox+2012}), and that the PN-ISM
  interaction is an interaction not directly between the expanding PN
  and the ISM, but rather between the PN and the previously set-up
  AGB-ISM bow shock.}  even for systems that have relatively low
speeds compared to the local ISM (i.e., space velocities, $v_{\rm{S}}
\lesssim 25$~$\kms$).  The presence of a well-defined bow shock
clearly separates interacting PNe from Str{\"o}mgren zones.  However,
the bow shock is not stable indefinitely.  At the later stages of the
PN-ISM interaction, instabilities, which develop over time in the
interaction region, begin to disrupt the bow-shock morphology.  The
disruption happens sooner, and is more severe, for the fastest-moving
systems.  For some systems, the disturbed PN might have a morphology
which resembles ionized, excited ISM.  Fortunately, the simulations
provide another distinguishing marker.  During the interaction,
material is ram-pressure stripped at the PN-ISM interface and
deposited downstream, forming a tail behind the PN.  Though diffuse,
tails have been observed behind three interacting PNe: Sh\,2-188
\citep{Wareing+2006sh188}, HFG\,1 \citep{Xilouris+1996,Boumis+2009},
and DeHt\,5 \citep{Ransom+2010}.  In the case of DeHt\,5, for which
bow-shock disruption is severe (see interaction classification of
\citealt{Ali+2012}), the ``thick'' tail stands as the best evidence of
a PN-ISM interaction.  In two other cases, namely Sh\,2-68 and
PHL\,932, faint downstream nebulosity could be an ionized trail in the
ambient ISM behind an old white dwarf \citep{Frew2008PhDT,Frew+2010}.

Recently, radio polarimetric observations have proven to be a good
tool for studying the environments of interacting PNe
\citep{Ransom+2008,Ransom+2010}.  Polarimetric observations at
frequencies $\leq$3~GHz (wavelengths $\geq$10~cm) are very sensitive
to Faraday rotation of the diffuse Galactic synchrotron emission
\citep[e.g.,][]{Landecker+2010}.  In propagating through an ionized
(and magnetized) medium, the polarized component of the background
emission is rotated at wavelength $\lambda$\,[m] through an angle
\begin{equation}
\Delta\theta = \rm{RM}\,\lambda^2\ [\rm{rad}],
\end{equation}
where RM is the rotation measure and depends on the line-of-sight
(LOS) component of the magnetic field, $B_{\|}$\,$[\mu\rm{G}]$, the
thermal electron density, $n_e$\,$[\rm{cm}^{-3}]$, and the path
length, $dl$\,$[\rm{pc}]$, as
\begin{equation}
\rm{RM} = 0.81\int{B_{\|}\,n_e\,dl\ [\rm{rad}\,\rm{m}^{-2}]}.
\end{equation}
Even in the relatively low electron-density environment of a PN tail,
a rotation at 1420~MHz ($\lambda = 21~\rm{cm}$) of the background
polarization angle of $\Delta\theta \gtrsim 10\arcdeg$ is possible,
and leads to a Faraday-rotation structure which is readily observable,
provided the background is relatively bright and smooth
\citep{Ransom+2010}.  In the interaction region, where the electron
density is higher and the ISM magnetic field likely compressed, the
resulting Faraday-rotation structure is even more conspicuous
\citep{Ransom+2008}.  Furthermore, a comparison of the Faraday
rotation through the leading and trailing structures may reveal a
deflection of the ISM magnetic field around the moving PN
\citep{Ransom+2008,Ransom+2010}.  If \sh\@ is indeed an interacting
PN, evidence in Faraday rotation of an interaction region, tail,
and/or deflected field should be ``visible'' in radio polarimetric
images.  If, on the other hand, \sh\@ is simply a Str{\"o}mgren-zone,
then these identifying features would be absent.  Naturally, any
ambient structures in the ISM around \sh\@, or elsewhere along the
LOS, could complicate the picture in either scenario.

In this paper, we present 1420~MHz radio images of the region around
\sh\@.  Our focus is linear polarization, but we also investigate
total intensity (Stokes-$I$) and neutral hydrogen (\ion{H}{1})
emission.  In Section~\ref{spacevel}, we look at the optical
morphology of \sh\@, and derive the space velocity of GD~561 relative
to the ISM.  In Section~\ref{obs}, we describe the preparation of the
radio images.  In Section~\ref{results}, we describe the polarization
(Faraday rotation), Stokes-$I$, and \ion{H}{1} structures which appear
in the images.  For the Faraday-rotation structures, we quantify
changes in the magnitude and/or sign of the RM from one structure to
the next.  In Section~\ref{discuss}, we discuss in detail the
structures observed in the radio images, and comment decisively on the
nature of \sh\@.  In Section~\ref{concl}, we give a summary of our
results and conclusions.

\section{THE OPTICAL MORPHOLOGY OF \sh\@ AND THE SPACE VELOCITY OF GD~561 \label{spacevel}}

\begin{figure}
\plotone{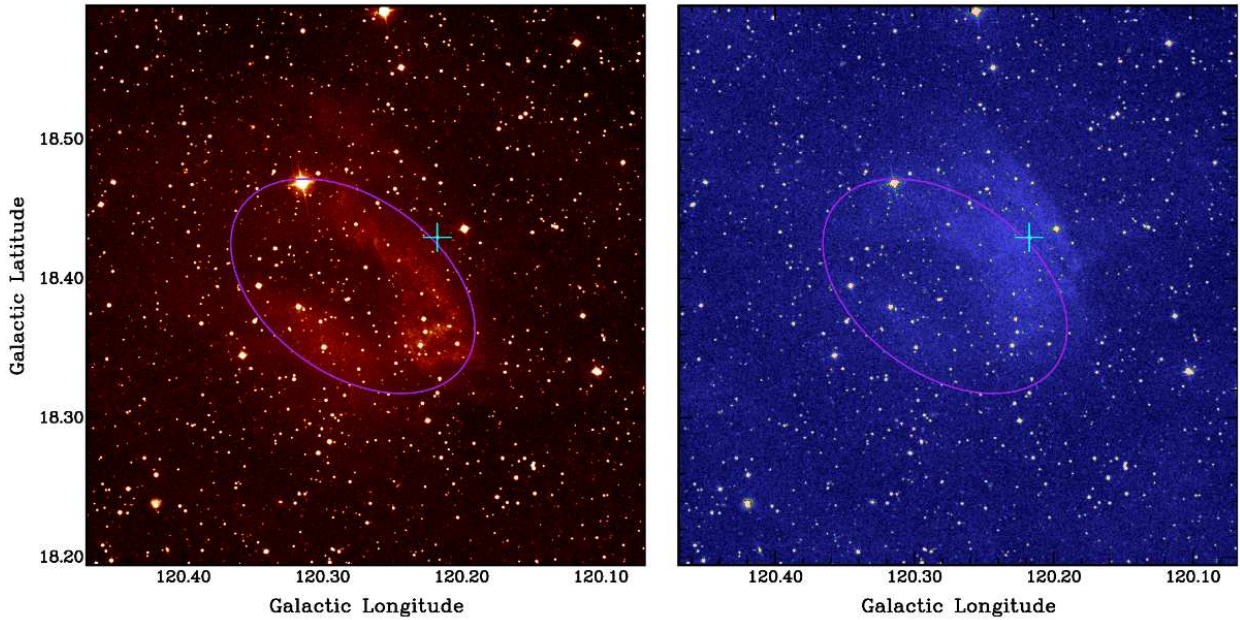}
\figcaption{R-band (left) and B-band (right) false-color DSS optical
  images of a $0.4\arcdeg$\,$\times$\,$0.4\arcdeg$ region around
  \sh\@.  Here and hereafter, images are presented in Galactic
  coordinates, with Galactic north up and Galactic east to the left.
  The intensity scale in each image is linear in photon counts, with
  lighter shades indicating higher counts.  The range of intensities
  in each image has been adjusted to show extended emission in the
  vicinity of \sh\@.  The angular resolution is $\sim$1\arcsec\@.  The
  elongated (magenta) ellipse drawn on each image, and on each
  subsequent image, shows the approximate extent of the R-band
  ``cleft-hoof'' structure.  The (blue) cross indicates the position
  of the white dwarf GD~561.
\label{DSSImageRedBlue}}
\end{figure}

We show in Figure~\ref{DSSImageRedBlue} R-band ($\lambda = 657$\,nm)
and B-band ($\lambda = 445$\,nm) Digitized Sky Survey (DSS) images for
a small $0.4\arcdeg$\,$\times$\,$0.4\arcdeg$ region around
\sh\@.\footnote{A detailed composite optical image of \sh\@ by
  T.\,A.\ Rector (University of Alaska Anchorage) and H.\ Schweiker
  (WIYN and NOAO/AURA/NSF) is available via the NOAO Image Gallery
  (\url{http://www.noao.edu/}).}  The R-band emission reveals the
so-called ``cleft-hoof'' morphology seen in $\rm{H}\alpha$ and
\ion{N}{2} \citep*{TweedyN1994}; i.e., an elongated shell-like
structure which is brightest in the (Galactic) southwest and fades
gradually to the northeast, where it blends with a ``halo'' of more
diffuse emission (see below).  We denote the cleft-hoof structure in
Figure~\ref{DSSImageRedBlue} with an ellipse having angular dimensions
$\sim$$12\arcmin$\,$\times$\,$\sim$$8\arcmin$.  Note that GD~561
(indicated by the cross) is located outside the cleft-hoof.  The
B-band emission, which traces \ion{O}{3} \citep*[see][]{TweedyN1994},
is approximately centered on GD~561.  Its intensity drops fairly
steadily east of GD~561, where it meets the cleft-hoof, but drops
rather sharply at the west edge.  The sharp edge is consistent with a
strong shock \citep*{TweedyN1994}, as might be expected for a
west-moving system (see below) in an interacting PN scanario.  On the
other hand, if the B-band edge really does signify a shock, then we
might expect an enhanced rim along the edge.  No such rim is present
in either the R-band or B-band.  We return to the discussion of the
nature of \sh\@ in Section~\ref{discuss} after presenting our radio
results.

\begin{figure}
\plotone{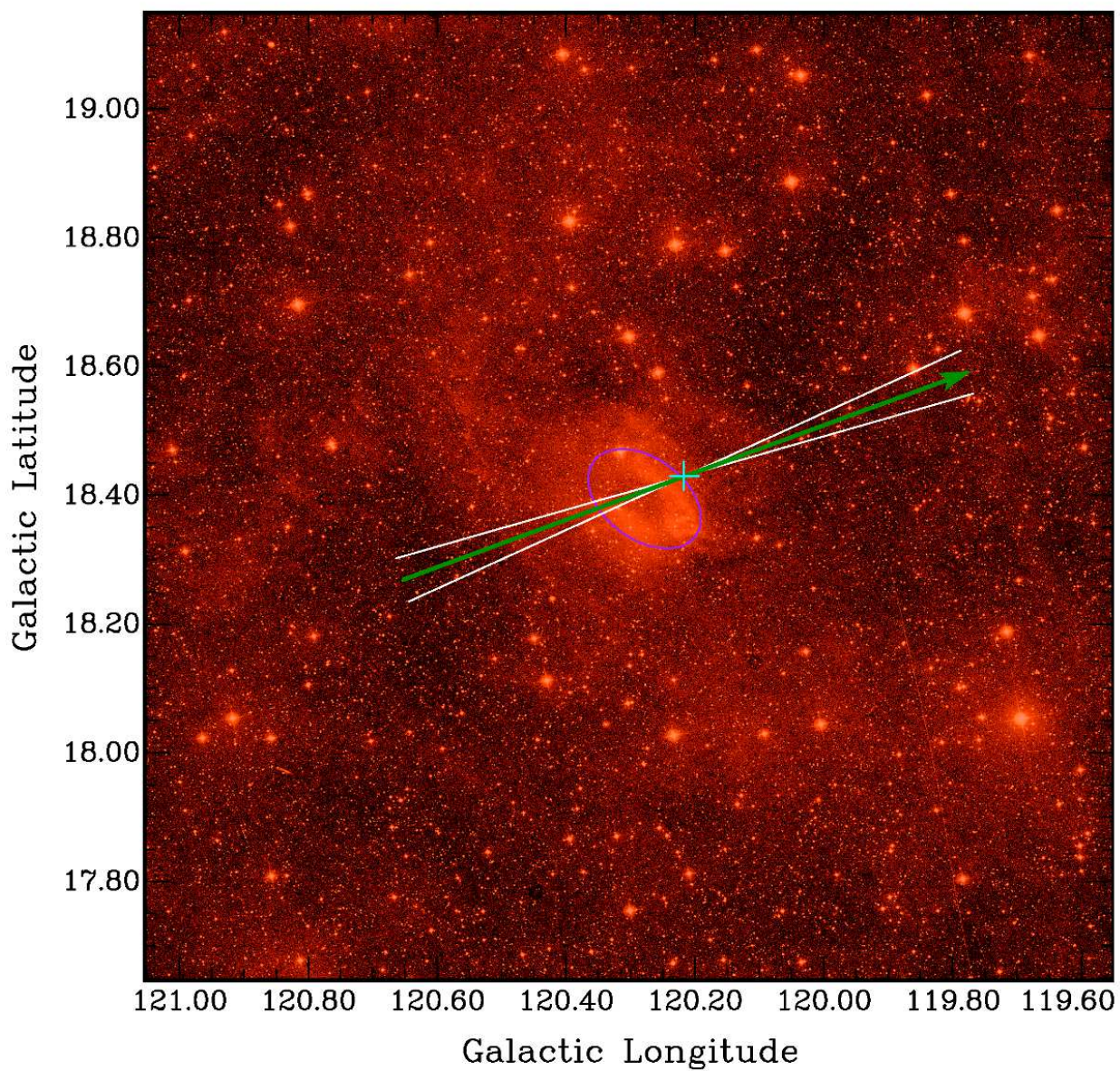}
\figcaption{R-band DSS image presented using a logarithmic intensity
  scale and zoomed out (compared to Fig.\,1) to show a
  $1.5\arcdeg$\,$\times$\,$1.5\arcdeg$ region around \sh\@.  The solid
  (dark green) arrow and adjacent (white) lines drawn on the image
  indicate the projected space velocity and space-velocity error cone
  of GD~561.  The length of the solid arrow represents the change in
  the sky position of GD~561 over $\sim$100,000~yrs, with its base
  starting $\sim$50,000~yr in the past and its tip ending
  $\sim$50,000~yr in the future.  Its present position is again
  indicated by the cross. The error cone represents largely the error
  range in the declination component of the proper motion.
\label{DSSImageRedLarge}}
\end{figure}

To provide a wider-field comparison with our radio images (see
Section~\ref{results}), we show in Figure~\ref{DSSImageRedLarge} the
R-band DSS image zoomed out to a $1.5\arcdeg$\,$\times$\,$1.5\arcdeg$
region around \sh\@.  We use a logarithmic scale here to highlight the
diffuse emission.  A low-surface-brightness halo surrounds the
brighter cleft-hoof structure.  Its extent is difficult to estimate,
since it fades into very faint ``background''\footnote{We use
  quotations to distinguish between emission observed on sightlines
  outside a particular structure of interest (``background'') and
  emission originating deeper along the LOS than the structure of
  interest (background).} structures.

The motion of GD~561 through the ambient ISM frames any discussion of
the nature of \sh\@, and for this reason we start by deriving the
star's space velocity.  The kinematic properties of GD~561 are
summarized in Table~\ref{thestar}.  Since no parallax has been
determined for GD~561, its distance, $\rm{d} = 415 \pm 120$~pc, is
estimated via its physical and photometric properties (also given in
Table~\ref{thestar}).  We use the position, distance, proper motion,
and radial velocity of GD~561 to estimate the UVW motion through the
ISM, where the U component is positive toward the Galactic center, V
is positive in the direction of Galactic rotation, and W is positive
toward the north Galactic pole \citep[e.g.,][]{JohnsonS1987}.  After
correcting for a solar motion of $(\rm{U}_{\sun}, \rm{V}_{\sun},
\rm{W}_{\sun}) = (11.10 \pm 1.25, 12.24 \pm 2.05, 7.25 \pm
0.62)$~$\kms$ \citep*{SchonrichBD2010}, we find velocity components
for GD~561 of $(\rm{U}, \rm{V}, \rm{W}) = (+59.5 \pm 7.1, +14.8 \pm
3.5, +18.4 \pm 3.9)$~$\kms$.\footnote{The uncertainty quoted for each
  of $\rm{U}_{\sun}$, $\rm{V}_{\sun}$, and $\rm{W}_{\sun}$ is the
  root-sum-square (rss) of the statistical standard error and the
  estimated systematic error for that component
  \citep[see][]{SchonrichBD2010}.  The uncertainties quoted for the
  space-velocity components of GD~561 were determined by computing
  10,000 draws of the component, each using different values for the
  input parameters (distance, proper motion, and radial velocity of
  GD~561, and $\rm{U}_{\sun}$, $\rm{V}_{\sun}$, and $\rm{W}_{\sun}$)
  chosen at random from within their error ranges, and then taking the
  standard deviation over the 10,000 draws.  The quoted uncertainties
  for the total, sky-projected, and LOS space velocities were
  determined in the same manner.} The total space velocity of GD~561
through the ambient ISM is then $v_{\rm{S}} = 64.1 \pm 8.1$~$\kms$,
with components on the sky and along the LOS of $63.2 \pm 8.2$~$\kms$
and $-10.4 \pm 1.2$~$\kms$, respectively.  Note, the relatively large
fractional uncertainties in the total and sky-projected space
velocities reflect largely the uncertainty in the distance estimate.
In contrast, the fractional uncertainty in the LOS component of the
space velocity reflects largely the uncertainty in $\rm{V}_{\sun}$.
The sky-projected space velocity of GD~561, including error cone, is
illustrated in Figure~\ref{DSSImageRedLarge}.

Our estimated $64.1 \pm 8.1$~$\kms$ space velocity for GD~561 is very
similar to the $66.2 \pm 5.8$~$\kms$ value computed recently for this
star by \citet{Ali+2013}.  We note, though, that \citet{Ali+2013} use
older values for the proper motion of GD~561, radial velocity of
GD~561, and assumed solar motion.  The small difference in the two
space-velocity estimates is thus somewhat coincidental.  Nevertheless,
based on its velocity relative to the ISM, \citet{Ali+2013} conclude
that GD~561 is a {\it probable} thin-disk star.  To more confidently
establish its membership as either a Galactic thin-disk or thick-disk
star, we would need a measurement of GD~561's parallax.

\section{OBSERVATIONS AND IMAGE PREPARATION \label{obs}}

The principal observing instrument for the data presented in this
paper is the Synthesis Telescope \citep[ST; see][]{Landecker+2000} at
the Dominion Radio Astrophysical Observatory (DRAO).  To image the
region around \sh\@, we used seven pointings (or {\it fields}) of the
ST.  We selected the fields so as to have one centered on the position
of \sh\@, and the remaining six form a hexagonal pattern around the
central field.  We chose to separate the field centers by $1\arcdeg$.
Since the ST antennas have a Gaussian primary-beam response at
1420~MHz with full-width-at-half-maximum (FWHM) of $107.2\arcmin$, the
chosen field separation yields a seven-field mosaic with a useful
diameter of $\sim$$3.8\arcdeg$.

The ST operates in the radio continuum at 408~MHz and 1420~MHz, and in
the 21~cm \ion{H}{1} line.  The 408~MHz continuum system receives only
right-hand circular polarization and is not used in this study.  The
1420~MHz continuum system has four 7.5 MHz bands centered on 1407.2,
1414.1, 1427.7, and 1434.6~MHz, respectively, and receives both
right-hand (R) and left-hand (L) circular polarizations.  The four
continuum bands are correlated separately, forming for each band a
full set of polarization products (RR, LL, RL, LR), and thus
recovering all four Stokes parameters ($I$, $Q$, $U$, $V$).  We
produced only $I$, $Q$, and $U$ mosaics, and show only those generated
by averaging the four bands.  The frequency corresponding to the
midpoint of the four bands is 1420.4~MHz.  The 5.0~MHz band around
this frequency is allocated to a 256-channel \ion{H}{1} spectrometer.
For this study we set the spectrometer to receive a total bandwidth of
1~MHz, providing a radial velocity range of 211~$\kms$ with a channel
separation of 0.824~$\kms$ and effective velocity resolution of
1.32~$\kms$.  The selected velocity range in the local standard of
rest (LSR) frame defined by \citet{SchonrichBD2010} is $-168\,\kms
\geq v_{\rm{LSR}} \leq +43\,\kms$.

The 1420~MHz continuum and \ion{H}{1} ST data were calibrated and
processed using standard procedures developed at DRAO
\citep[e.g.,][]{Taylor+2003,Landecker+2010}.  We note that the
wide-field instrumental polarization (i.e., ``leakage'' from $I$ into
$Q$ and $U$) for the seven fields presented in this paper was
calibrated using the visibility-plane technique developed by
\citet{Reid+2008} rather than the older image-plane technique
described in \citet{Taylor+2003}.  In producing our seven-field $Q$
and $U$ mosaics, we cut off each individual field at a radial
separation $\rho = 75\arcmin$ from the field center, since the
root-mean-square (rms) residual instrumental polarization error
becomes larger than $1\%$ outside this radius. (For point of
comparison, we use a more liberal cut-off in producing the $I$ and
\ion{H}{1} mosaics, namely $\rho = 93\arcmin$.) The rms instrumental
polarization error is further reduced in the mosaicing process.

The noise level in the central $\sim$$2.5\arcdeg$ of our ST mosaics is
approximately uniform due to the arranged overlap between the central
field and the outer six fields.  In our continuum (band-averaged) $I$,
$Q$, and $U$ mosaics, the estimated rms noise level in the central
region is $\sim$0.021~K ($\sim$0.16~$\rm{mJy}\,\rm{beam}^{-1}$).  Some
artifacts are seen above this level in the immediate vicinity of the
brightest point sources.  In our \ion{H}{1} mosaics the estimated
noise level (per channel) in the central region is $\sim$2~K rms.
Outside the central region the noise level in each mosaic increases
due to the corrected primary-beam response in the outer six fields.

The ST is sensitive at 1420~MHz to emission from structures with
angular sizes of $\sim$45\arcmin\@ down to the resolution limit of
$\sim$1\arcmin\@.  To recover structures on the largest spatial
scales, we added to our $I$ mosaic single-antenna data from the
Stockert-25m \citep{Reich1982} and Effelsberg-100m \citep{Reich+2004}
surveys and to our $Q$ and $U$ mosaics single-antenna data from the
DRAO-26m\footnote{The DRAO-26m telescope has recently been named the
  John Galt Telescope.} \citep{Wolleben+2006} and Effelsberg-100m
surveys. (We did not add single-antenna data to our \ion{H}{1}
mosaics.)  Striping is present at $\sim$3 times the rms noise level in
the low-resolution Stockert-25m and DRAO-26m surveys for the region
around \sh\@, but filtered out in the first stage of the merging
process (i.e., when the data are merged with the Effelsberg-100m
data).  A full description of the process by which the datasets are
combined is given in \citet{Taylor+2003}.  The effective noise levels
in the resulting (final) images are unchanged from those noted above
for the ST mosaics.

\section{RADIO IMAGES OF THE REGION AROUND \sh\@ \label{results}}

\subsection{1420~MHz Polarization \label{PolImages}}

\begin{figure}
\plotone{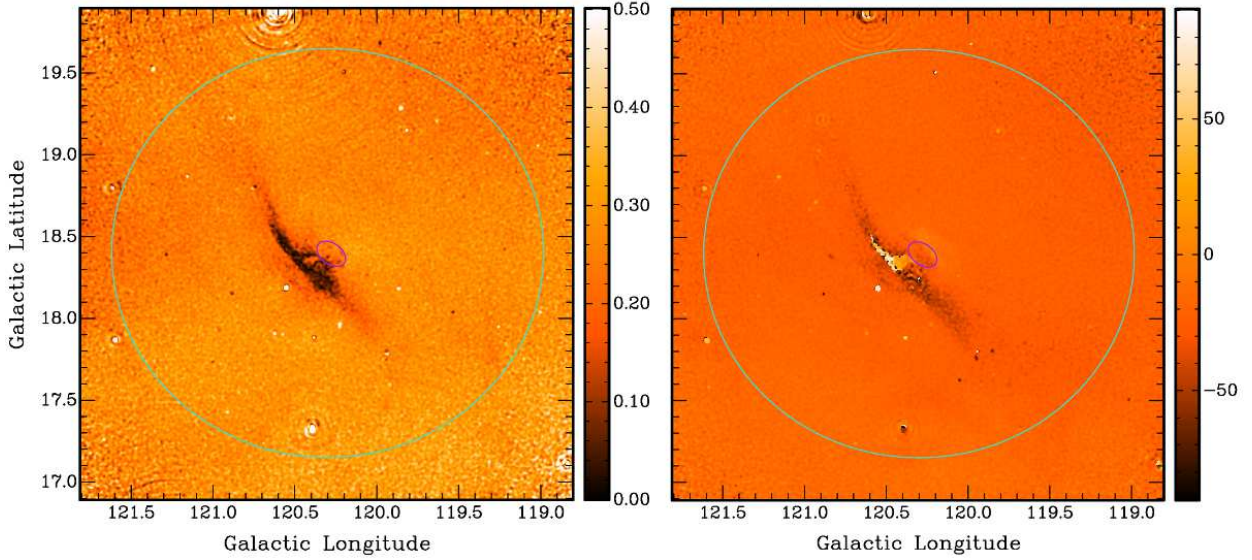}
\figcaption{1420~MHz polarized intensity (left) and polarization angle
  (right) images of a $3\arcdeg$\,$\times$\,$3\arcdeg$ region around
  \sh\@.  The color choice is arbitrary.  The polarized-intensity
  scale is linear in brightness temperature, with lighter shades
  indicating higher temperatures.  The range of intensities has been
  cut off at 0.500~K to highlight the diffuse emission.  The
  polarization-angle scale is linear and ranges from $-90\arcdeg$
  (black) to $+90\arcdeg$ (white).  Note that abrupt black-to-white
  transitions do not represent large changes in angle, since
  polarization angles of $-90\arcdeg$ and $+90\arcdeg$ are equivalent
  (i.e., polarization angles are modulo $180\arcdeg$).  The solid
  $2.5\arcdeg$-diameter (turquoise) circle marks the edge of our
  central field.  The rms noise is approximately uniform inside this
  circle.  The resolving beam at the center of each image is
  $1.03\arcmin$\,$\times$\,$0.99\arcmin$ (FWHM), oriented at a
  position angle (east of north) of $-2\arcdeg$.  The resolving beam
  varies in size over each image by $\sim$2\%.
\label{PolImagesLarge}}
\end{figure}

In Figure~\ref{PolImagesLarge} we show the polarized intensity ($P =
\sqrt{Q^2+U^2-(1.2\sigma)^2}$; e.g., \citealt{SimmonsS1985}, where the
last term gives explicitly the noise bias correction) and polarization
angle ($\theta_P = \frac{1}{2}\arctan{U/Q}$) images at 1420~MHz for
the $3\arcdeg$\,$\times$\,$3\arcdeg$ region around \sh\@.
Polarization angles are measured east of north.  The small
$\sim$$12\arcmin$\,$\times$\,$\sim$$8\arcmin$ ellipse denotes (as in
Figures~1 and 2) the approximate extent of the R-band nebula.  The
large $2.5\arcdeg$-diameter circle outlines the central field, within
which the rms noise is approximately constant (see Section~\ref{obs}).
The images reveal a thin structure next to \sh\@ which stands out, due
to Faraday rotation, in stark contrast to the relatively smooth
surroundings: the polarized intensity is significantly lower than that
of the surroundings, and the polarization angles show large variations
and have a mean value significantly different from that of the
surroundings. (We describe the mechanism and effects of the Faraday
rotation in detail below.)  The structure is aligned approximately
northeast-southwest, with its northern tip bending slightly
north-northeast and its southern tip bending slightly south-southwest.
At its midpoint, this $S$-shaped structure (for lack of a better
descriptor) stretches westward into a ``bridge'' which appears to
terminate at the eastern edge of \sh\@.  The images also reveal
lower-contrast Faraday-rotation features which extend east and west of
the midpoint of the $S$-shaped structure.

\begin{figure}
\plotone{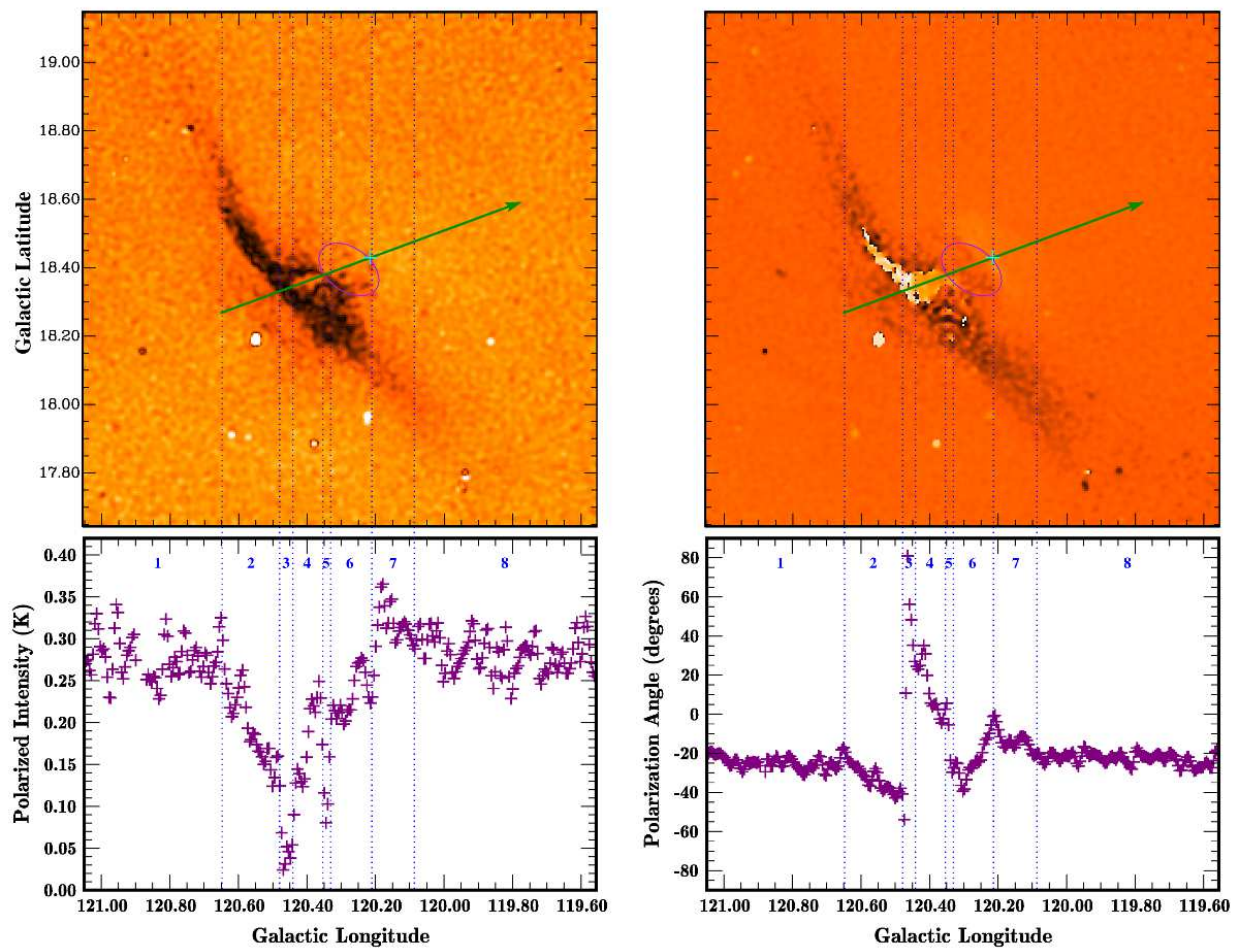}
\figcaption{Top panels: The polarized intensity (left) and
  polarization angle (right) images from Fig.\,3 zoomed in to show the
  same $1.5\arcdeg$\,$\times$\,$1.5\arcdeg$ region presented in
  Fig.\,2.  The projected space velocity and current position of
  GD~561 are indicated (see Fig.\,2 caption).  Bottom panels:
  Polarized intensities (left) and polarization angles (right) along
  the projected trajectory of GD~561.  The (blue) dotted lines show
  the east-west extent of the eight slice-regions described in the
  text.
\label{PolImagesWithSlices}}
\end{figure}

To see more detail in the Faraday-rotation structures around \sh\@, we
show in Figure~\ref{PolImagesWithSlices} (top panels) the central
$1.5\arcdeg\times1.5\arcdeg$ region of our polarized intensity and
polarization angle images.  We include (as in
Figure~\ref{DSSImageRedLarge}) a vector indicating the projected space
velocity of GD~561, as well as a marker for the current position of
GD~561.  Not coincidentally (see Section~\ref{discuss}), some of the
more distinct boundaries and systematic variations in the
Faraday-rotation structures can be traced along the projected space
velocity vector.  To accentuate this fact, we present in
Figure~\ref{PolImagesWithSlices} (bottom panels) polarized intensity
and polarization angle slices along this line.  We have identified
eight slice-regions whose edges represent a clear boundary or a change
in the pattern of the variations.  To summarize, briefly, regions 1
and 8 are the ``surroundings.'' Region 2 is the first indication in
the images (moving from east-to-west) of a change from the
surroundings.  Region 3 marks the narrow eastern boundary with the
$S$-shaped structure.  Region 4 corresponds to the bridge that
connects the $S$-shaped structure to \sh\@.  Region 5 marks the narrow
eastern (or downstream) boundary of \sh\@.  Region 6 runs across
\sh\@, from the downstream boundary to approximately the position of
GD~561.  Finally, region 7 runs across the relatively wide western (or
upstream) boundary of \sh\@.  In Section~\ref{sliceregions}, we
describe the actions of Faraday rotation in each of our slice-regions.
In Section~\ref{Sshape}, we look at the Faraday-rotation behavior
along the major axis of the $S$-shaped structure.  In
Section~\ref{RManalysis}, we describe the analysis which yields
estimates of the RMs through the various structures.

\subsubsection{Faraday rotation in structures along the projected trajectory of GD~561 \label{sliceregions}}

We describe here the Faraday rotation in the eight identified
slice-regions along the projected space velocity vector of GD~561 (see
Figure~\ref{PolImagesWithSlices}).  We begin with the surroundings
(regions 1 and 8), which set comparables for the observed polarized
intensities and polarization angles, and then proceed westward through
regions 2--7.  Some mention is made here of the physical conditions
responsible for the Faraday rotation, but we save a more complete
discussion for Section~\ref{discuss}.

Regions 1 and 8: The polarized intensities and polarization angles in
these slice-regions are approximately constant within the noise.
Moreover, the mean values for region 1 are very similar to those for
region 8: $0.271 \pm 0.026$~K and $-24\arcdeg \pm 3\arcdeg$ in region
1, and $0.275 \pm 0.023$~K and $-22\arcdeg \pm 2\arcdeg$ in region 8.
These slice values do not differ (within the quoted standard
deviations) from mean values taken over larger, two-dimensional,
versions of regions 1 and 8 having the same longitude boundaries.  In
other words, the slices are fair representations of the greater
surroundings.

Region 2: Moving east-to-west across this region, we see the polarized
intensity decrease from the mean region 1 value, and the polarization
angle rotate systematically clockwise (i.e., to more negative values)
compared to the mean region 1 value.  To assist the reader in
interpreting these variations, we give here a description of the
mechanism through which Faraday rotation affects the observed
intensities and angles.  The specifics apply to region 2, but the
principles apply to each of regions 2--7.  The mechanism is as
follows: diffuse linearly-polarized background passes through region
2, which is ionized and magnetized, with one or more of the electron
density, magnitude of the LOS magnetic field, and path length
increasing east-to-west.  Equivalently, we can say there is a RM
gradient across region 2.  The background is thus Faraday rotated (see
Equation~1), with the degree of Faraday rotation increasing with more
westward sightlines.\footnote{Note, we use rotation both in the
  Faraday rotation sense, i.e., the systematic change in the
  background polarization angle along a particular sightline, and to
  describe the systematic change in observed polarization angles over
  a range of sightlines.  We believe that the intended use is clear in
  each instance.} The rotated background is then (vector) added to the
foreground to produce the observed values for the polarized intensity
and polarization angle.  Not surprisingly, the polarization angles for
region 2 differ from those of region 1, since region 1 results from
the addition of the unrotated background and the foreground.  Also,
the polarized intensities for region 2 are lower than those in region
1, since the rotated background is not as well aligned with the
foreground as is the unrotated background.  This effect is referred to
as localized depth depolarization or background-foreground
cancellation \citep[e.g.,][]{Gray+1999,Uyaniker+2003}.  Beam
depolarization, due to systematic variations or stochastic
fluctuations in the polarization angle on scales smaller than the beam
size ($\sim$$1\arcmin$ for the ST), can also reduce intensities (e.g.,
\citealt{Gray+1999}; \citealt{Uyaniker+2003};
\citealt*{HaverkornKd2004a}).  Via the technique described in
Section~\ref{RManalysis}, we estimate a mean RM in region 2 of $-6 \pm
2$\,$\rm{rad}\,\rm{m}^{-2}$, and find that beam depolarization
accounts for $\sim$$30\%$ of the intensity reduction.  (Note, the
relative contribution of beam depolarization is higher in the
low-intensity ``knots'' seen outside the slice but within the same
longitude boundaries.) Since the systematic variation in polarization
angle observed in region 2 amounts to only $\sim$$1.5\arcdeg$ over a
beamwidth, the estimated $\sim$$30\%$ beam depolarization must arise
from stochastic fluctuations.  The most likely cause of such
fluctuations is turbulence.

Region 3: This narrow region is marked by near-zero polarized
intensities and rapid variations in the polarization angles.  Moving
east-to-west, the polarization angles first appear to jump
pseudo-randomly from negative to positive values, and then settle into
a rapid clockwise rotation.  It may be, however, that the rotation is
systematically clockwise throughout region 3, with the rotation being
extremely rapid over the first three data points.  This interpretation
is believable to the eye, especially if we ignore the second point.
Indeed, the $\sim$$+11\arcdeg$ value for this point can easily be
explained as an east-to-west average, over a beamwidth, of values near
$-90\arcdeg$ and values near $+90\arcdeg$.  Moreover, a systematic
clockwise rotation across region 3 is consistent with the change in
mean RM between regions 2 and 4.  In any case, the result of the rapid
polarization-angle variations is strong beam depolarization across
region 3.  We do not estimate a mean RM for this region.

Region 4: Moving east-to-west across this region, we see a general
increase in polarized intensity and a systematic clockwise rotation of
polarization angle.  These trends demonstrate that the rotated
background aligns better with the foreground for western sightlines
than for eastern sightlines.  We estimate a mean RM in this region of
$-44 \pm 5$\,$\rm{rad}\,\rm{m}^{-2}$.  This is the largest absolute RM
value found for any of the observed Faraday-rotation structures,
indicating that region 4 has the highest combination of electron
density and LOS magnetic field anywhere within $\sim$$1.5\arcdeg$ of
\sh\@.

Region 5: This narrow region is marked by a sharp drop in polarized
intensity and an apparent discontinuity in polarization angle.  The
discontinuity is supported, and indeed magnified, by a comparison of
the mean RMs found for regions 4 and 6.  What looks like a
$\sim$$-20\arcdeg$ (east-to-west) change in polarization angle between
the second and third data points, is more likely a $\sim$$+160\arcdeg$
change.  For the intensities, however, it is the effective
polarization-angle change, modulo $180\arcdeg$, that contributes to
background-foreground cancellation and beam depolarization.  So, while
the drop in intensity across region 5 is significant, it is not as
severe as that seen in region 3.  We do not estimate a mean RM for
this region.

Region 6: Moving east-to-west across this region, we see, after the
first five points, a systematic counter-clockwise rotation in
polarization angle.  The polarized intensity is approximately constant
for the first five points, increases to a peak value just west of the
midpoint, and finally drops to a local minimum at the transition with
region 7.  Not surprisingly, the intensity peak is coincident with
polarization angles which have approximately the same value as the
mean surroundings; i.e., background-foreground cancellation is a
minimum at this location.  The drop in intensity at the transition
between regions 6 and 7 is due to a combination of
background-foreground cancellation and beam depolarization.  The
systematic east-to-west change in polarization angle from values more
negative than those for the mean surroundings to values more positive
than those for the mean surroundings indicates a change in the sign of
the RM over region 6.  Physically, this change demonstrates a reversal
in the direction of the LOS component of the magnetic field moving
east-to-west across region 6.  We estimate a mean RM in this region of
approximately zero ($\rm{RM} = -1 \pm 2$\,$\rm{rad}\,\rm{m}^{-2}$).

Region 7: Moving east-to-west across this region, we see the
polarization angle rotate clockwise, level off, and then continue
rotating clockwise to the transition with region 8.  Interestingly,
the eastern half of region 7 has a mean polarized intensity which is
slightly higher than that of the surroundings.  There are two possible
explanations for this: First, there could be an enhancement of the
diffuse polarized emission somewhere along these sightlines.  This is
very unlikely, given that these enhanced intensities span an angle of
just $\sim$$0.1\arcdeg$.  Second, instead of background-foreground
cancellation, we could have for these sightlines background-foreground
enhancement.  This is possible if the background was rotated in the
eastern half of region 7 so as to align better with the foreground
than it does in the surroundings.  This interpretation is consistent
with the average background and foreground properties we estimate in
Section~\ref{RManalysis}.  We estimate a mean RM in this region of $+8
\pm 2$\,$\rm{rad}\,\rm{m}^{-2}$.

\subsubsection{Faraday rotation along the $S$-shaped structure \label{Sshape}}

The $S$-shaped structure is the largest contiguous Faraday-rotation
feature in the observed region around \sh\@.  It runs
$\sim$1.2$\arcdeg$ from top to bottom, and varies slightly in width
from $\sim$0.1$\arcdeg$ for the north ``arm'' to $\sim$0.15$\arcdeg$
for the bridge and south ``arm''.  Internally, the $S$-shaped
structure contains several notable substructures.  Along with the
bridge, the most eye-catching is the jagged-edged,
$\sim$0.3$\arcdeg$-long ``finger'' seen most easily in polarization
angle.  We encountered the finger in Section~\ref{sliceregions} as
region 3, the eastern boundary of the central bridge.  In two
dimensions, we see that the finger extends from the bridge into the
north arm.  The jagged outline of the finger is $\sim$1$\arcmin$ wide
and characterized by near-zero polarized intensities and variations in
polarization angle of $>$70$\arcdeg$ (see slice data for region 3).

Unlike the north arm's finger, the south arm has no discernible
extended substructures.  Immediately southwest of the bridge, we see
1$\arcmin$--2$\arcmin$-wide fluctuations in both polarized intensity
and polarization angle.  Moving farther away from the bridge on the
south arm, we continue to see small-scale fluctuations, but with
maximum deviations decreasing with angular separation from the bridge.
A similar trend is seen in the small-scale fluctuations beyond the
finger in the north arm.  Averaging over 3--4 fluctuations a
separation $\sim$$0.3\arcdeg$ from the center of the bridge, we
estimate for each arm $\rm{RM} = -16 \pm 5$\,$\rm{rad}\,\rm{m}^{-2}$.
Over the same area, we also estimate that beam depolarization accounts
for $\sim$$50\%$ of the intensity reduction.  The visible fluctuations
and high degree of beam depolarization point to considerable
turbulence in the arms of the $S$-shaped structure.

\subsubsection{RM analysis \label{RManalysis}}

In the absence of a foreground, and assuming the background is uniform
across the observed region, the RM in a localized structure can be
estimated directly by taking the difference in polarization angle
between sightlines on the structure and sightlines on the
surroundings.  When the foreground cannot be ignored, as for the
Faraday-rotation structures around \sh\@, the scenario is more
complex.  Nevertheless, it is possible, through trial and error, to
construct a model that resolves the differences in observed polarized
intensity and polarization angle between the surroundings and each
structure of interest \citep[e.g.,][]{WollebenR2004,Sun+2007}.  The
model parameters include for each structure the background intensity
and angle, the foreground intensity and angle, the RM in the
structure, and the degree of beam depolarization on the structure.  As
in the simple case for a negligible foreground, the model assumes that
the background (and now also the foreground), are the same for
sightlines on the surroundings and each structure of interest.  We
believe this to be a fair assumption for the
$1.5\arcdeg\times1.5\arcdeg$ region around \sh\@, since, outside the
Faraday-rotation structures themselves, the observed intensities and
angles are approximately constant.  Using this trial-and-error
technique, we were able to find an acceptable range of background and
foreground values, and estimate the RM and degree of beam
depolarization for each of regions 2, 4, 6, and 7 along the projected
trajectory of GD~561, as well as for the arms of the $S$-shaped
structure.  We did not construct models for regions 3 and 5, since
these regions show rapid variations in intensity and angle over a
small number of pixels.  Our model background has polarized intensity
$0.14 \pm 0.02$~K and polarization angle $-30\arcdeg \pm 10\arcdeg$,
while our model foreground has polarized intensity $0.13 \pm 0.02$~K
and polarization angle $-15\arcdeg \pm 10\arcdeg$ (see also
Table~\ref{RMTable}).  Our estimated mean RMs are given in the
appropriate sections above and presented in Table~\ref{RMTable}.  Our
estimated beam depolarization values are also presented in
Table~\ref{RMTable}.  We note that, while we present only the mean
RMs, the relatively smooth variations in polarization angle across
each modeled region demonstrate that the RMs also vary smoothly across
each region.  We note, also, that the quoted RM uncertainties are
dominated by the rms variations in the observed polarization angles,
and not by the uncertainties in the model background and foreground.
In fact, because the background and foreground must add to produce the
surroundings, the estimated RMs are relatively insensitive to even
larger (forced) changes to either the background or foreground.
Nonetheless, it is reassuring that Galactic three-dimensional emission
models \citep{Sun+2008,SunR2009} also produce for this part of the sky
a background ($d \gtrsim 415$~pc) and foreground ($d \lesssim 415$~pc)
with polarized-intensity ratio $\sim$1 and polarization-angle
difference $\sim$10--15$\arcdeg$.

RMs estimated via modeling are subject to ambiguity.  For most
sightlines around \sh\@, we adopt the smallest possible rotation,
since the observed polarization angles vary by $\lesssim$$20\arcdeg$
as we move from the surroundings to the structure of interest.  The
exception is region 4.  The smallest mean RM through region 4 is
$\rm{RM} = +26 \pm 5$~$\rm{rad}\,\rm{m}^{-2}$.  This option is
inconsistent with the transition in polarization angle from
$-90\arcdeg$ (black) to $+90\arcdeg$ (white) at the region's east,
north, and south boundaries (see Figure~\ref{PolImagesWithSlices}).
The next smallest mean RM for region 4 is $\rm{RM} = -44 \pm
5$~$\rm{rad}\,\rm{m}^{-2}$.  A negative RM more naturally explains the
observed transitions.

Since the ST has four bands, we looked for point sources that could be
used to probe the RMs on sightlines through regions 2--7 and the arms
of the $S$-shaped structure \citep*[e.g.,][]{SavageSF2013}.
Unfortunately, no point sources are so fortuitously located.  However,
it is also possible to estimate RMs directly from the diffuse
emission, provided the polarization angles in each band have
relatively small rms scatter.  We performed a four-band analysis (see
\citealt*{BrownTJ2003}; \citealt{Geisbuesch+2014}) on the diffuse
emission at the midpoints of each of regions 2, 4, 6, and 7, and found
RM values consistent with the mean values estimated using the modeling
method: a region-by-region comparison shows that the sense of rotation
determined by each method is the same and that the magnitude
determined by each method differs by no more than the combined errors.
The four-band result for region 4 ($\rm{RM} = -39 \pm
7$~$\rm{rad}\,\rm{m}^{-2}$) validates the argument given above in
favor of a negative RM for region 4.

\subsection{1420~MHz Total Intensity \label{TiImage}}

\begin{figure}
\plotone{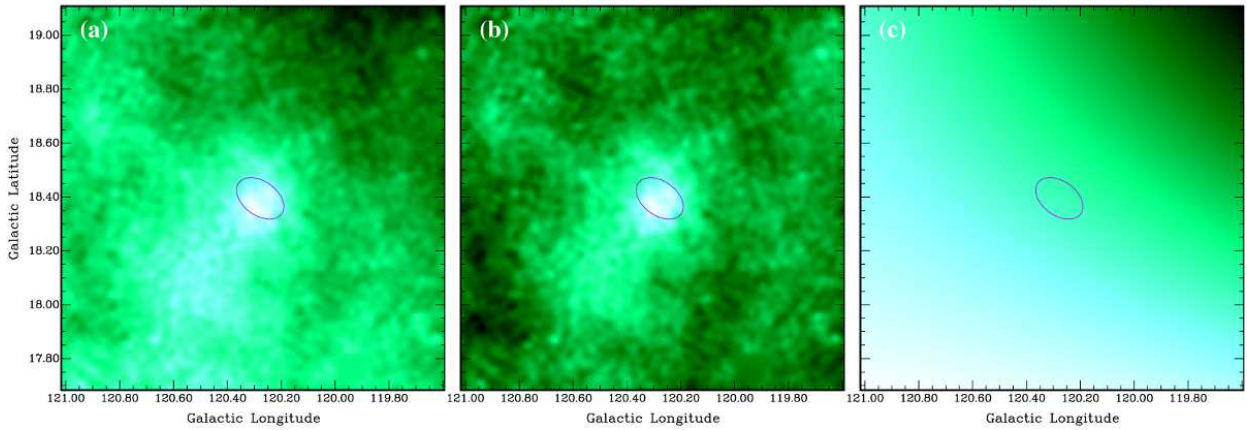}
\figcaption{1420 MHz total intensity ($I$) image of a
  $1.4\arcdeg$\,$\times$\,$1.4\arcdeg$ region around \sh\@ (panel
  (a)).  Point sources are removed and the image smoothed to
  $2\arcmin$.  Panel (b) is the $I$ image presented in panel (a) after
  subtraction of a twisted plane (panel (c)).  The color choice is
  arbitrary.  The intensity scale in each panel is linear in
  brightness temperature, with lighter shades indicating higher
  temperatures.  The intensities range before subtraction from 4.264~K
  to 4.515~K (panel (a)) and after subtraction from $-0.065$~K to
  0.142~K (panel (b)).  The intensities in the twisted plane vary
  smoothly from 4.284~K to 4.424~K (panel (c)).
\label{TiImages}}
\end{figure}

\begin{figure}
\plotone{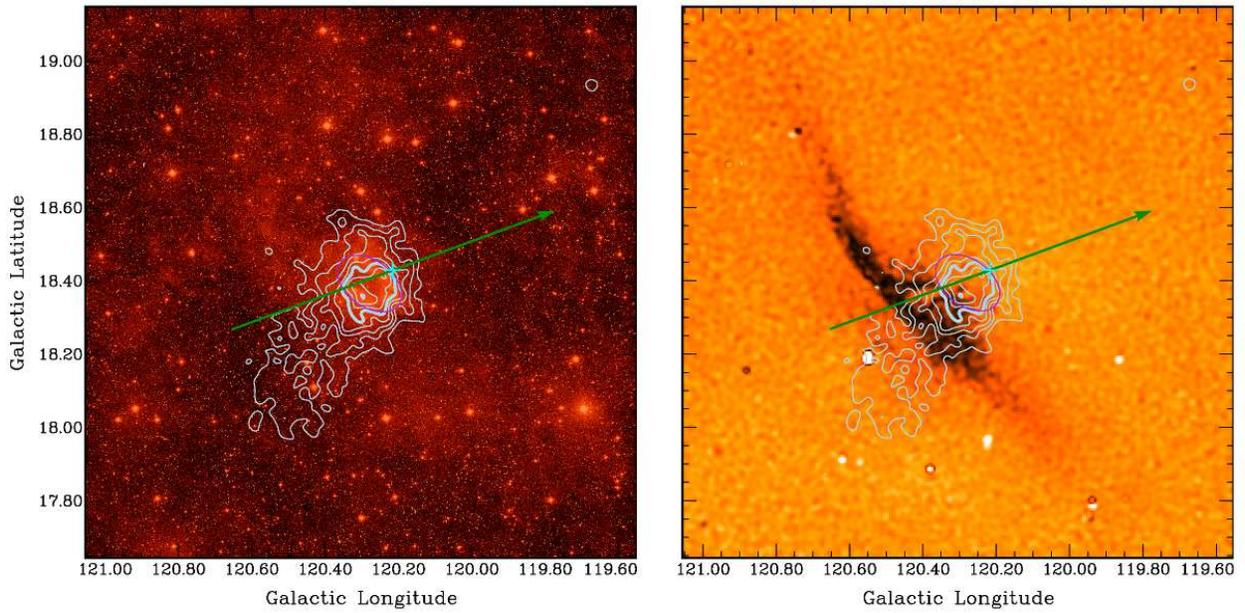}
\figcaption{DSS R-band (left) and polarized intensity (right) images
  from Figs.\,2 and 4, respectively, with 1420~MHz $I$ contours
  overlaid (in light blue).  The contours were generated from the
  image presented in Fig.\,5b.  Contour levels increase by 0.014~K
  (the estimated rms ``off-source'' variations) starting at a
  brightness temperature of 0.042~K.  The 0.098~K contour is drawn
  thicker than the remaining contours.  The projected space velocity
  and current position of GD~561 are indicated (see Fig.\,2 caption).
\label{TiContours}}
\end{figure}

In Figure~\ref{TiImages}a we show the total intensity ($I$) image for
the region around \sh\@, after removing point sources and smoothing
the resulting image (modestly) to $2\arcmin$.  In
Figure~\ref{TiImages}b we show the $I$ image after also subtracting a
twisted plane representative of the diffuse ``background''
(Figure~\ref{TiImages}c).\footnote{The twisted plane was fitted to the
  image data using a least-squares algorithm, and subtracted using the
  program {\tt fittwist} from the DRAO Export Package
  \citep*{HiggsHW1997}.  A twisted plane is a surface in which the
  gradient vector ``twists'' about an axis perpendicular to the image
  plane.  Similar routines are standard parts of many packages for
  analysis of radio astronomical data.}  The mean residual brightness
temperature (in Figure~\ref{TiImages}b) away from the emission
surrounding \sh\@ is $<$0.01~K, with rms variations 0.014~K.  In
Figure~\ref{TiContours} we superimpose $I$ contours (generated from
the image presented in Figure~\ref{TiImages}b) on the DSS R-band and
1420~MHz polarized intensity images for the region around \sh\@.  The
brightest $I$ emission (i.e., that within the thick 0.098~K contour)
is coincident spatially with \sh\@.  Moreover, the localized
$\sim$0.14~K and $\sim$0.13~K peaks align precisely (i.e., within
$\sim$$1\arcmin$) with the mid-lines of the southeast and northwest
``rims'' of the R-band cleft-hoof structure.  The $I$ emission on and
near these peaks almost certainly shares the same (thermal) origin as
the R-band emission. (No enhancement is seen in the 408~MHz emission
at or near the position of \sh\@, which further supports a thermal
origin for the 1420~MHz $I$ emission.) In contrast, the brightest $I$
emission has no overall counterpart in Faraday rotation.  This is
perhaps not surprising, since thermal emission requires only
relatively high electron densities, while Faraday rotation also
requires a relatively strong LOS magnetic field.

At brightnesses just below $\sim$0.10~K, the $I$ emission forms a
$\sim$$0.4\arcdeg$-diameter halo around \sh\@.  A halo is also present
in the R-band emission.  Curiously, at even lower brightnesses
($\sim$0.04 to $\sim$0.06~K), the $I$ emission stretches southeast
into an extended ``plateau'' $\sim$$0.4\arcdeg$ long and
$\sim$$0.3\arcdeg$ wide.  No similarly elongated structure is seen in
either R-band emission (above the DSS ``background'') or Faraday
rotation.  However, the plateau does pass directly across the bridge
of the $S$-shaped structure.

\subsection{Neutral Hydrogen (\ion{H}{1}) \label{HiImages}}

\begin{figure}
\plotone{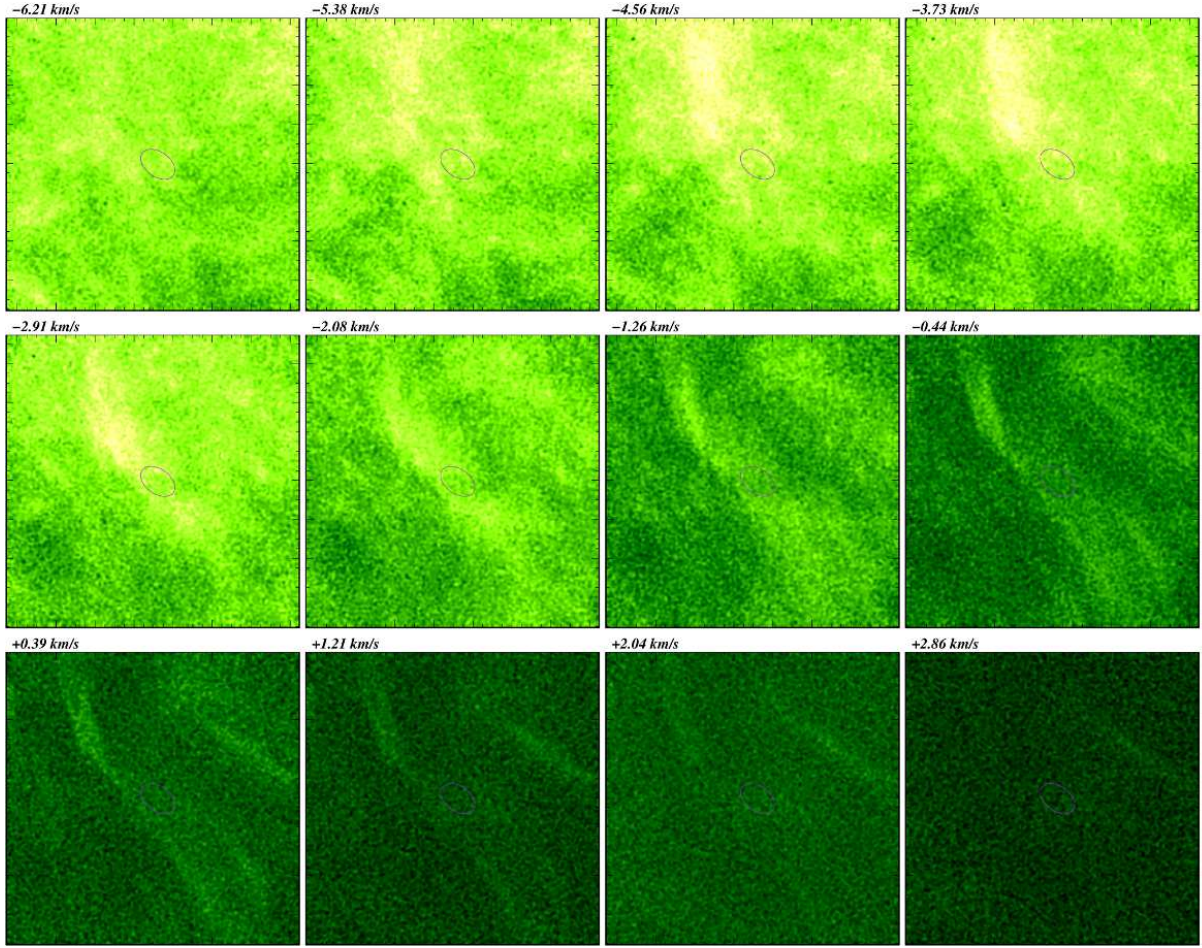}
\figcaption{Continuum-subtracted \ion{H}{1} images of a
  $1.5\arcdeg$\,$\times$\,$1.5\arcdeg$ region around \sh\@
  region. (The displayed region is the same as that presented in
  Fig.\,2 and Fig.\,4.)  The twelve displayed velocity channels are a
  subset of the 256 total channels.  The velocity with respect to the
  \citet{SchonrichBD2010} LSR is indicated for each image.  The color
  choice is arbitrary.  The intensity scale is linear in brightness
  temperature, with lighter shades indicating higher brightness
  temperatures.
\label{TwelveHiImages}}
\end{figure}

In Figure~\ref{TwelveHiImages} we show \ion{H}{1} images of the region
around \sh\@ for twelve consecutive velocity channels between
$v_{\rm{LSR}} = -6.21$~$\kms$ and $v_{\rm{LSR}} = +2.86$~$\kms$.  The
central 8--10 channels in this range contain emission from two local
(i.e., $v_{\rm{LSR}} \sim 0$~$\kms$) \ion{H}{1} clouds -- one which
runs approximately northeast-to-southwest through the center of the
displayed region, and one which is confined more to the northwest
corner of the region.  The full spatial extent of these clouds is best
seen in channels which are otherwise essentially empty (i.e.,
$v_{\rm{LSR}} = -0.43\,\kms$, $+0.39\,\kms$, $+1.22\,\kms$).  The
cloud in the northwest corner is of no particular interest here.  The
more elongated, central cloud, however, has the same position,
orientation, and profile as the $S$-shaped, Faraday-rotation
structure.

\begin{figure}
\plotone{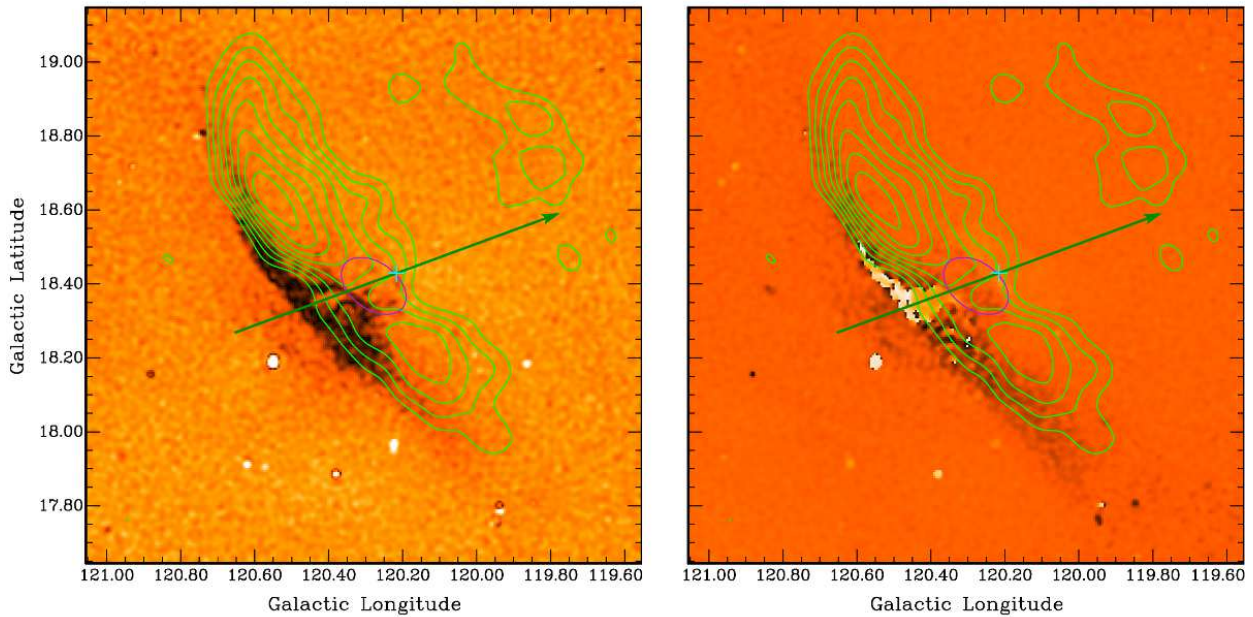}
\figcaption{Polarized intensity (left) and polarization angle (right)
  images from Fig.\,4 with \ion{H}{1} contours overlaid (in green).
  The contours were generated after averaging the eight central
  velocity channels ($v_{\rm{LSR}} = -4.56$~$\kms$ to $v_{\rm{LSR}} =
  +1.21$~$\kms$) presented in Fig.\,7, subtracting a twisted plane,
  and smoothing the resulting \ion{H}{1} image to $5\arcmin$.  Contour
  levels increase by 0.7~K (the estimated rms ``off-source''
  variations) starting at a brightness temperature of 2.1~K.  The
  projected space velocity and current position of GD~561 are
  indicated (see Fig.\,2 caption).
\label{HiContours}}
\end{figure}

To showcase the correspondence between the elongated \ion{H}{1} cloud
and the Faraday-rotation structures, we superimpose in
Figure~\ref{HiContours} \ion{H}{1} contours on the 1420~MHz polarized
intensity and polarization angle images.  The \ion{H}{1} contours were
generated after averaging the eight central channels ($v_{\rm{LSR}} =
-4.56$~$\kms$ to $v_{\rm{LSR}} = +1.21$~$\kms$), subtracting a twisted
plane representative of the ``background,'' and smoothing the
resulting image to $5\arcmin$.  The mean residual brightness
temperature away from the cloud (after subtraction of the twisted
plane) is $<$0.5~K, with rms variations 0.7~K.  The superposition
demonstrates clearly that the \sh\@-centered position of the
\ion{H}{1} cloud is not a LOS coincidence.  The neutral material in
this cloud and the ionized material in the Faraday-rotation structures
(and \sh\@) are physically interconnected: The arms of the $S$-shaped
structure run precisely along the sharp eastern edge of the cloud.
Moreover, the finger (seen best in the polarization angle image),
which stretches north from the midpoint of the $S$-shaped structure,
actually fills a slight indentation in the eastern edge of the cloud.
Moving west from this boundary, along the projected trajectory of
GD~561, we see the ionized material (in the Faraday-rotation bridge)
``puncture'' the neutral material.  \sh\@ itself sits in a
low-surface-brightness cavity at approximately the midpoint of the
cloud.

\begin{figure}
\plotone{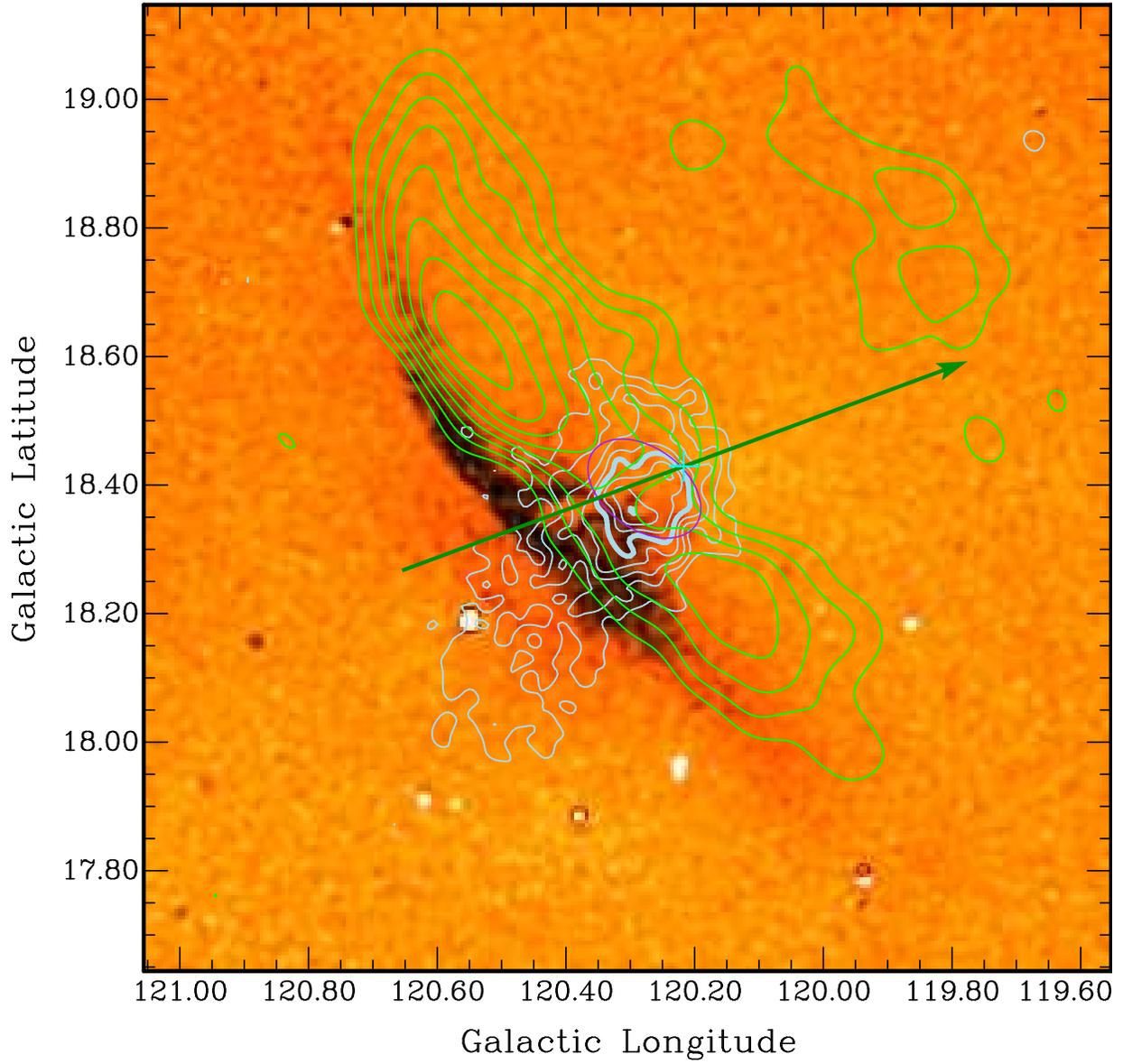}
\figcaption{Polarized intensity image from Fig.\,4 with both $I$
  (Fig.\,6) and \ion{H}{1} (Fig.\,8) contours overlaid.  The projected
  space velocity and current position of GD~561 are indicated (see
  Fig.\,2 caption).
\label{AllInOne}}
\end{figure}

In Figure~\ref{AllInOne} we superimpose both the \ion{H}{1} and $I$
contours on the 1420~MHz polarized intensity image, yielding an {\it
  all-in-one} radio image.  We noted in Section~\ref{TiImage} that
there is no overall correspondence between the $I$ emission and the
Faraday-rotation structures seen in polarization, but that the
extended $I$-emission plateau passes across the bridge of the
$S$-shaped structure.  We see now, in comparing the contours, that the
low-surface-brightness $I$ emission is shaped by the \ion{H}{1} cloud:
the $I$ halo is indented northeast and southwest of \sh\@ at precisely
the locations where the \ion{H}{1} emission is brightest, while the
$I$ plateau extends southeast through the depression (including
\sh\@-filled cavity) between the \ion{H}{1}-emission peaks.  We
discuss the likely origins for the halo and plateau in
Section~\ref{StromgrenInterpretation} and
Section~\ref{OriginOfPlateau}, respectively.

\section{DISCUSSION \label{discuss}}

\subsection{Electron Density in \sh\@ \label{PNElectronDensity}}

We can derive the emission measure (EM) for \sh\@ from the thermal $I$
emission coincident with the R-band cleft-hoof structure
\citep*[see][]{WilsonRH2012}.  The mean brightness temperature over
the structure is $T_b = 0.110 \pm 0.005$~K.  The localized southeast
and northwest peaks have brightness temperatures $T_b = 0.142 \pm
0.014$~K and $T_b = 0.129 \pm 0.014$~K, respectively. (We use for the
uncertainty in the peak brightness temperatures the rms ``off-source''
variations, and for the mean brightness temperature the peak-to-peak
variations in the estimate over reasonable changes in the position of
the structure's boundary.  All subsequent peak/region
brightness-temperature uncertainties are estimated in the same way.)
The electron temperature has not been directly measured for \sh\@
(e.g., from emission-line intensity ratios).  We can, however, adopt
$T_e = 10000 \pm 2000$~K, which reasonably covers the range of
electron temperatures found for \ion{H}{2} regions and PNe
\citep*[e.g.,][]{NichollsDS2012}.  We then derive $\rm{EM} = 70 \pm
6$~$\rm{cm}^{-6}$\,pc for the mean value over the cleft-hoof
structure, and $\rm{EM} = 91 \pm 9$~$\rm{cm}^{-6}$\,pc and $\rm{EM} =
82 \pm 11$~$\rm{cm}^{-6}$\,pc for the southeast and northwest peaks,
respectively.

To turn the EMs into electron densities ($\rm{EM} =
\int{{n_e}^2\,dl}$), we need to estimate the path length through the
emission regions.  The on-sky dimensions of the cleft-hoof structure
are $\sim$$12\arcmin$\,$\times$\,$\sim$$8\arcmin$.  For the average
angle-equivalent path length, we estimate $8\arcmin \pm 2\arcmin$
(where the nominal $8\arcmin$ corresponds to the average path length
through a $12\arcmin$-diameter sphere.) At $\rm{d} = 415 \pm 120$~pc,
this translates to an average physical path length $\Delta l = 0.97
\pm 0.37$~pc.  Since the localized peaks lie at the rims of the
structure, their angle-equivalent path lengths are likely less than
$8\arcmin$.  We estimate an angle-equivalent path length through each
peak of $6\arcmin \pm 2\arcmin$, which translates to a physical path
length of $\Delta l = 0.72 \pm 0.32$~pc.  Thus we derive a mean
electron density over \sh\@ $n_e = 8 \pm 2$~$\rm{cm}^{-3}$, and
electron densities for each peak $n_e = 11 \pm 3$~$\rm{cm}^{-3}$.

\subsection{Hydrogen Density in the \ion{H}{1} Cloud \label{HiDensity}}

Our radio observations show clearly that there is an elongated
\ion{H}{1} cloud intrinsic to the region around \sh\@.  The brightness
temperatures in this cloud are relatively low for compact \ion{H}{1}
structures \citep[see, e.g.,][]{Stil+2006}, even at the emission
peaks, suggesting that the cloud is optically thin.  If this is the
case, we can derive the neutral hydrogen column density ($N_H$) of the
cloud directly from the measured brightness temperatures
\citep[see][]{WilsonRH2012}.  The mean (velocity-averaged) brightness
temperature over the $\sim$$1.2\arcdeg$-long and
$\sim$$0.4\arcdeg$-wide cloud is $T_b = 3.5 \pm 0.5$~K.  The
$5\arcmin$-smoothed peaks northeast and southwest of \sh\@ have
brightness temperatures $T_b = 7.5 \pm 0.7$~K and $T_b = 4.9 \pm
0.7$~K, respectively.  With a velocity-width for the cloud $\Delta \nu
= 6.6 \pm 0.4$~$\kms$, we derive $N_H = (4.2 \pm 0.7) \times
10^{19}$~$\rm{cm}^{-2}$ for the mean column density over the cloud,
and $N_H = (9.1 \pm 1.1) \times 10^{19}$~$\rm{cm}^{-2}$ and $N_H =
(5.9 \pm 1.0) \times 10^{19}$~$\rm{cm}^{-2}$ for the northeast and
southwest peaks, respectively.\footnote{As noted in
  Section~\ref{HiImages}, the \ion{H}{1} cloud is seen in 8--10
  velocity channels.  The presented brightness temperatures correspond
  to the 8-channel average.  If we instead average ten channels, the
  brightness-temperature estimates decrease by $\sim$20\%.  The
  estimated column densities, which depend on the product of $T_b$ and
  $\Delta \nu$, do not change.}

We now need to estimate the path length through the emission regions.
Since the aspect ratio of the \ion{H}{1} cloud is $\gtrsim$3:1
(allowing for the very-low-surface-brightness emission at the
northernmost and southernmost ends), and we have no independent
information on the LOS dimension, any adopted path length will be
quite uncertain.  That said, the \ion{H}{1} contours (see Figures~8,
9) give the impression that the cloud could be a fragment of an old,
large shell.  If this is the case, the \ion{H}{1} cloud is more like a
sheet than a filament, with the $\sim$$1.2\arcdeg$ long (on-sky) axis
more closely approximating the path length than the $\sim$$0.4\arcdeg$
short axis.  For the average angle-equivalent path length through the
cloud, we therefore estimate $1.0\arcdeg \pm 0.4\arcdeg$.  This
translates, at $\rm{d} = 415 \pm 120$~pc, to a physical path length
$\Delta l = 7.2 \pm 3.6$~pc.  Given the $\sim$$50\%$ uncertainty in
this estimate, we adopt the same path length for the peaks.  Thus we
derive a mean hydrogen density over the cloud $n_H = 1.9 \pm
1.0$~$\rm{cm}^{-3}$, and hydrogen densities $n_H = 4.0 \pm
2.1$~$\rm{cm}^{-3}$ and $n_H = 2.6 \pm 1.4$~$\rm{cm}^{-3}$ for the
(smoothed) northeast and southwest peaks, respectively. (We show two
significant figures here to allow an inter-comparison of the mean and
peak values.) For point of interest, we note that there is CO emission
at $v_{\rm{LSR}} \sim 0$~$\kms$ coincident with the denser northeast
peak \citep{Heithausen+1993}.

\subsection{ISM Magnetic Field Around \sh\@ \label{ISMMagneticField}}

Our measured RMs sample the LOS component of the ISM magnetic field
through the observed Faraday-rotation structures around \sh\@.  The
signs of the RMs are negative throughout the $S$-shaped structure, and
for regions 2--4 along the projected trajectory of GD~561, but change
systematically across region 6 from negative to positive, and remain
positive across region 7.  Negative RMs require a magnetic field with
LOS component pointing into the plane of the sky, while positive RMs
require a field with LOS component pointing out-of the plane of the
sky.  Thus, our results demonstrate that the LOS component of the ISM
field changes from predominantly into-the-sky for regions east of
\sh\@ to out-of-the-sky for the region immediately west of \sh\@.
Since the RMs vary systematically (and smoothly) across the face of
\sh\@ (region 6), the transition in LOS field direction from
into-the-sky to out-of-the-sky is likely also smooth.  This points to
a localized deflection of the ambient ISM field around \sh\@ rather
than an abrupt, intrinsic change.

If we can estimate the electron density (or, equivalently, the EM) in
and path length through the various regions, we can use our estimated
RMs to derive via Equation~2 the corresponding magnitudes for the LOS
field in each region ($B_{\|} \approx \rm{RM} /
(0.81\sqrt{\rm{EM}\,\Delta l})$).  Regions 4 and 7 fall within the
extended $I$-emission halo around \sh\@.  Though the halo's LOS
dimension may be larger than that for the Faraday-rotation regions, we
assume the mean EMs on sightlines through the $I$ emission apply also
to these regions.  The mean brightness temperature is $T_b = 0.070 \pm
0.005$~K for region 4 and $T_b = 0.060 \pm 0.005$~K for region 7.
Assuming $T_e = 10000 \pm 2000$~K in these ionized regions, we
estimate a mean EM of $45 \pm 5$~$\rm{cm}^{-6}$\,pc in region 4 and
$38 \pm 5$~$\rm{cm}^{-6}$\,pc in region 7.  For the angle-equivalent
path lengths, we estimate for each region $0.15\arcdeg \pm
0.05\arcdeg$.  These path lengths represent on the low end the
approximate widths of the Faraday-rotation region and on the high end
the path length through a spherical $\sim$$0.32\arcdeg$-diameter-halo
along sightlines $\sim$$0.05\arcdeg$ from its edge.  At $\rm{d} = 415
\pm 120$~pc, the angle-equivalent path length for each region
translates to a physical path length $\Delta l = 1.1 \pm 0.5$~pc.  For
region 4, with $\rm{RM} = -44 \pm 5$\,$\rm{rad}\,\rm{m}^{-2}$, we
therefore derive for the (mean) LOS field $B_{\|} = 8 \pm
2$~$\mu\rm{G}$.  For region 7, with $\rm{RM} = +8 \pm
2$\,$\rm{rad}\,\rm{m}^{-2}$, we derive $B_{\|} = 1.5 \pm
0.5$~$\mu\rm{G}$, with opposite direction of course.

The magnitude difference between region 4 and region 7 likely does not
reflect a change in the strength of the ambient field, but rather a
change from a field that has a significant LOS component in region 4
to a (deflected) field that lies largely in the plane of the sky in
region 7.  If we assume that the ambient field within the \ion{H}{1}
cloud lies as much in the plane of the sky as along the LOS, and that
its LOS component is well-approximated by our region 4 value, the
intrinsic strength of the cloud's field is $11 \pm 3$~$\mu\rm{G}$.
This is consistent with the statistically-determined mean field
strength for diffuse \ion{H}{1} sheets and filaments
\citep[$\sim$10~$\mu\rm{G}$, see][]{HeilesT2005}, but $\gtrsim$2 times
the average azimuthal field in the general ISM
\citep[$\sim$4.2~$\mu\rm{G}$, see][]{Heiles1996b}.  Observations of
the Zeeman splitting of the 21~cm line might better constrain the
estimates of the LOS magnetic field in the \ion{H}{1} cloud, and would
not be limited to sightlines through the Faraday-rotation structures.

\subsection{A Str{\"o}mgren-Zone Interpretation for \sh\@ \label{StromgrenInterpretation}}

Recently, various authors have argued that \sh\@ in not an authentic
PN, but rather ambient ISM material ionized by GD~561
\citep*[e.g.,][]{FrewP2010}.  What do our radio observations have to
say about this interpretation?  Thermal radio emission has not
previously been detected for \sh\@.  Our 1420~MHz $I$ contours reveal
emission coincident with the R-band cleft-hoof structure, as well as
emission from a $\sim$$0.4\arcdeg$-diameter halo around \sh\@.  The
radial distribution of the emission, starting from the center of \sh\@
and moving outward to the edge of the halo, gives the impression of
two separate emission structures; namely, a relatively
high-surface-brightness structure with radius $\sim$$0.1\arcdeg$
(\sh\@), and a relatively low-surface-brightness structure with radius
$\sim$$0.2\arcdeg$ (halo).  If the emission arose instead from one,
approximately spherical, structure, we would expect the decrease in
intensity to be more gradual.  Our estimated densities also point to
\sh\@ being a separate structure: the mean electron density over \sh\@
($n_e = 8 \pm 2$~$\rm{cm}^{-3}$) is a factor $\sim$2 larger than the
peak hydrogen density in the ambient \ion{H}{1} cloud ($n_H = 4 \pm
2$~$\rm{cm}^{-3}$).  It is possible, however, that GD~561 has ionized
a relatively dense region of ambient material near the center of the
cloud.

Our 1420~MHz polarization images reveal Faraday-rotation structures in
the region(s) surrounding GD~561, with the most prominent structures
appearing downstream.  Faraday rotation is expected in the
Str{\"o}mgren zone around an isolated white dwarf
\citep[see][]{Ignace2014}.  Also, since a white dwarf is expected to
leave behind an ionized trail \citep*[see][]{McCulloughB2001}, it may
be reasonable to find downstream Faraday rotation as well.  In this
particular case, though, an ionized trail is inconsistent with the
observations.  First, the downstream Faraday rotation isn't subtle,
which indicates a significant enhancement in electron density (or
magnetic field).  If GD~561 is moving through the warm ionized ISM,
with ambient ionization fraction $\gtrsim$0.8 \cite[see,
  e.g.,][]{Sembach+2000}, the enhancement in an ionized trail relative
to the surroundings would be marginal.  Second, the eastward extent of
the Faraday rotation is too small.  The first indication of Faraday
rotation, moving east-to-west along the projected trajectory of
GD~561, occurs at $l = 120.65\arcdeg$ (see
Figure~\ref{PolImagesWithSlices}).  At the projected space velocity of
GD~561, this corresponds to $\sim$50,000~yr.  Even assuming a
relatively dense ISM outside the cloud ($n_e \sim n_H \sim
1$~$\rm{cm}^{-3}$), the recombination time in an ionized trail is
$t_{\rm{rec}} \approx (n_e \alpha_{H})^{-1} = 120,000$~yr.\footnote{We
  assume $T_{\rm{gas}} \approx 10^4$~K for the ionized trail, and thus
  use a recombination rate coefficient $\alpha_{H} = 2.59 \times
  10^{-13}$~$\rm{cm}^{-3}\,\rm{s}^{-1}$ \citep*{OsterbrockF2006}.}
Finally, the $S$-shaped Faraday-rotation structure along the east edge
of the cloud represents a region of greatly enhanced electron density
(and likely magnetic field).  Since the material in an ionized trail
is ambient ISM, there is no reason for it to be {\it building-up}
outside the cloud.

The observed Faraday rotation presents an additional difficulty for
the Str{\"o}mgren-zone interpretation.  Changes in RM along the
projected trajectory of GD~561 indicate that the ISM magnetic field is
significantly deflected around \sh\@.  Such a deflection cannot easily
be explained by a Str{\"o}mgren zone.  The presence and motion of
GD~561 alone should not alter the ambient field.  GD~561 could have a
residual wind, but it seems unlikely that such a wind would be dense
enough to produce the observed deflection.

Even if \sh\@ itself is not a Str{\"o}mgren zone, the $I$-emission
halo surrounding \sh\@ likely does represent the extent to which
GD~561's ionizing photons penetrate the ambient \ion{H}{1} cloud.
Using the range of effective temperatures (see Table~\ref{thestar})
and the estimated bolometric magnitude \citep[see][]{Frew2008PhDT} for
GD~561, we estimate that ionizing photons are emitted at a rate
$L^{\star}_{\rm{91.2}} = (3.5 \pm 1.9) \times
10^{45}$~$\rm{photons}\,\rm{s}^{-1}$.  We adopt a (mean) angular
radius for the halo of $\sim$$0.2\arcdeg$, which, at $\rm{d} = 415 \pm
120$~pc, translates to a physical radius $R_{S} = 1.45 \pm 0.42$~pc.
With this estimated luminosity and radius, we derive (assuming
$T_{\rm{gas}} \approx 10^4$~K for the halo) a Str{\"o}mgren-zone
particle density $n_e = n_p = 6^{+4}_{-3}$~$\rm{cm}^{-3}$.  This
particle density is consistent with what we find for the neutral
hydrogen density near the peaks in the \ion{H}{1} cloud, and suggests
that the uninterrupted cloud had, rather than two dense knots, a
modest north-to-south density gradient along its central ridge.  Since
the density at the edges of the cloud are lower, we see the halo
perimeter stretch out on either side of the ridge (see
Figure~\ref{AllInOne}).  Along position angles approximately
orthogonal to the ridge, ionizing photons from GD~561 likely escape
the cloud altogether.

\subsection{A PN-ISM Interpretation for \sh\@ \label{PNISMInterpretation}}

\begin{figure}
\centerline{
\includegraphics[bb = 0 2 460 400,width=0.45\textwidth,clip]{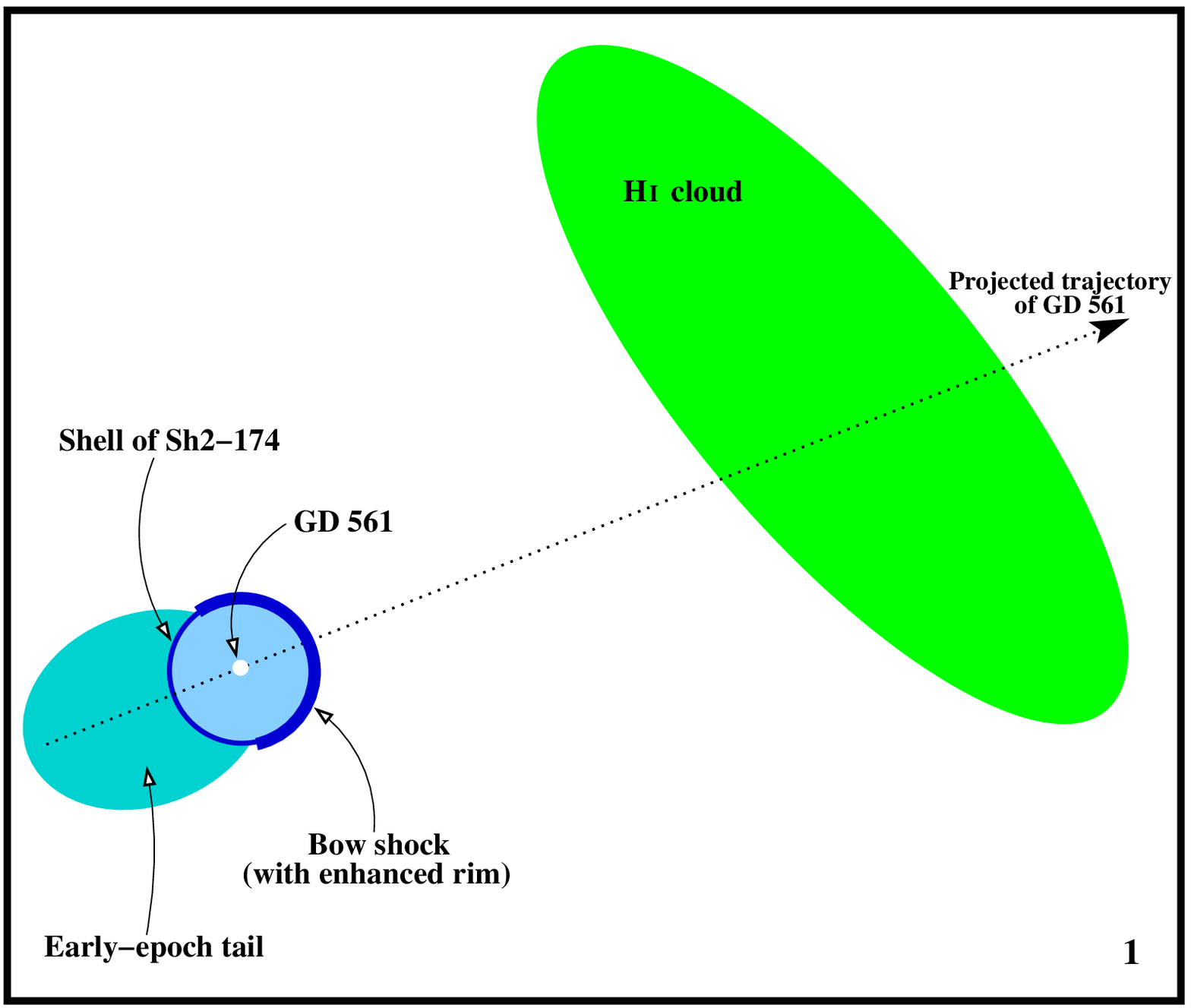}
\includegraphics[bb = 0 2 460 400,width=0.45\textwidth,clip]{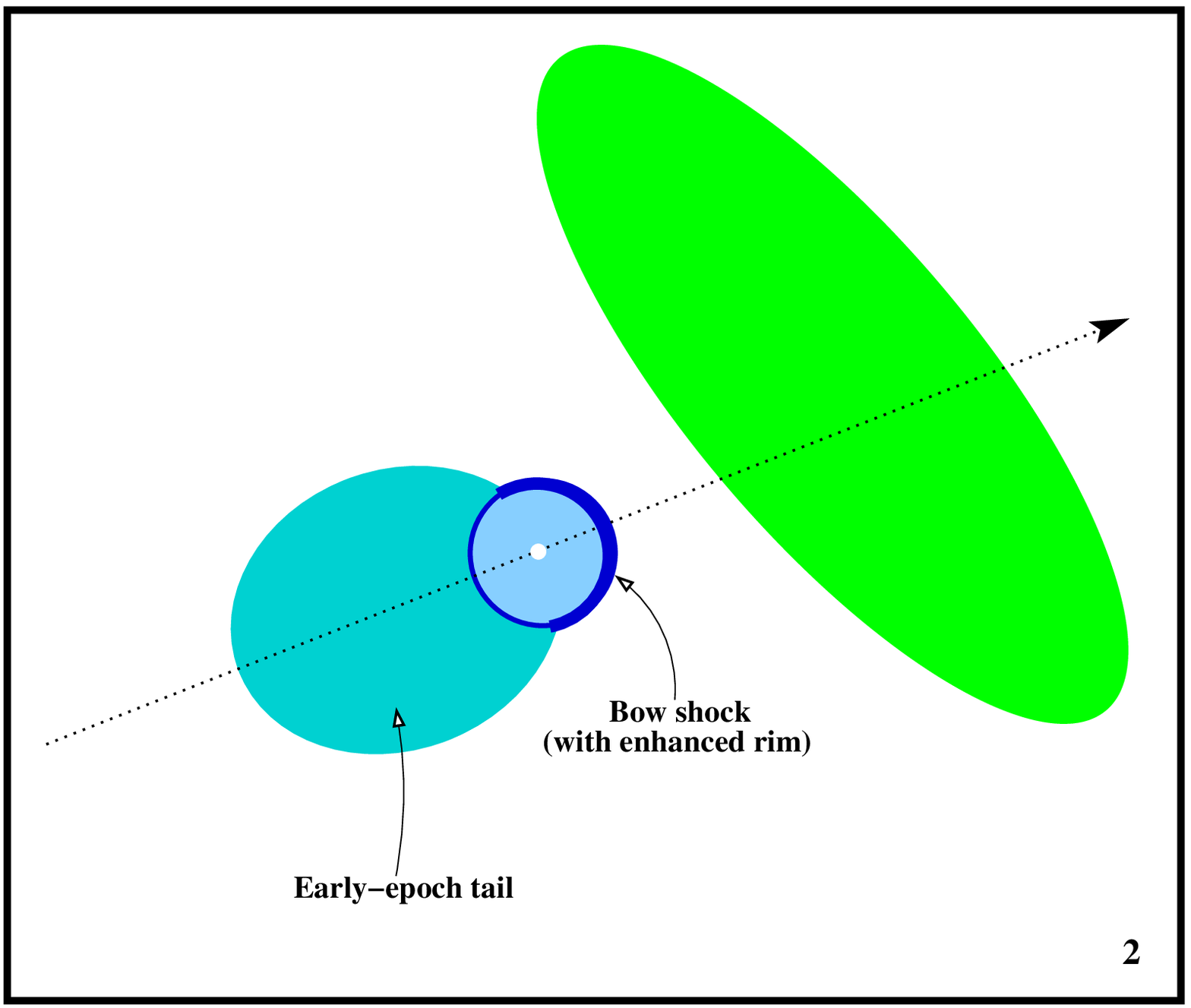}}
\centerline{
\includegraphics[bb = 0 2 460 400,width=0.45\textwidth,clip]{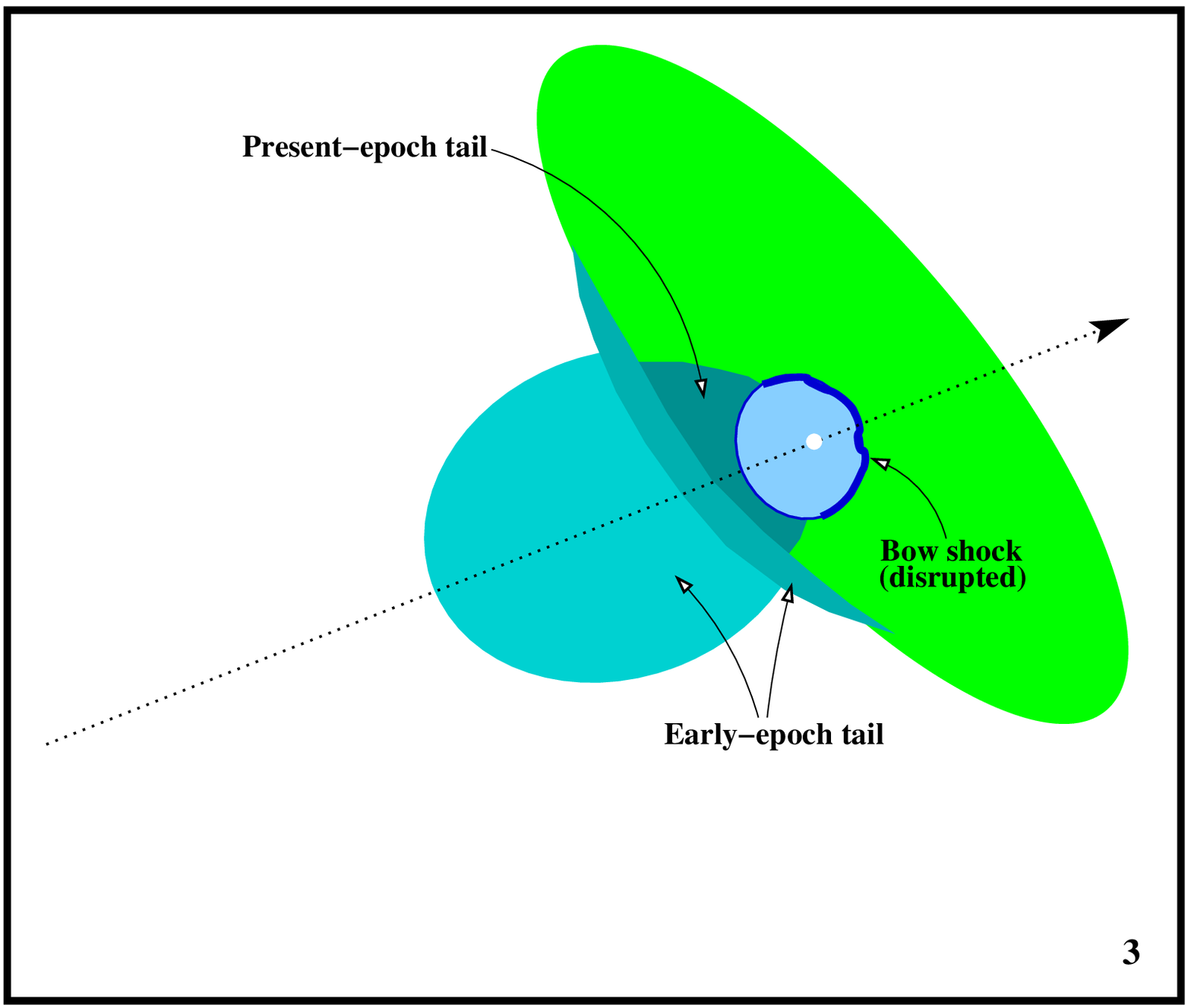}
\includegraphics[bb = 0 2 460 400,width=0.45\textwidth,clip]{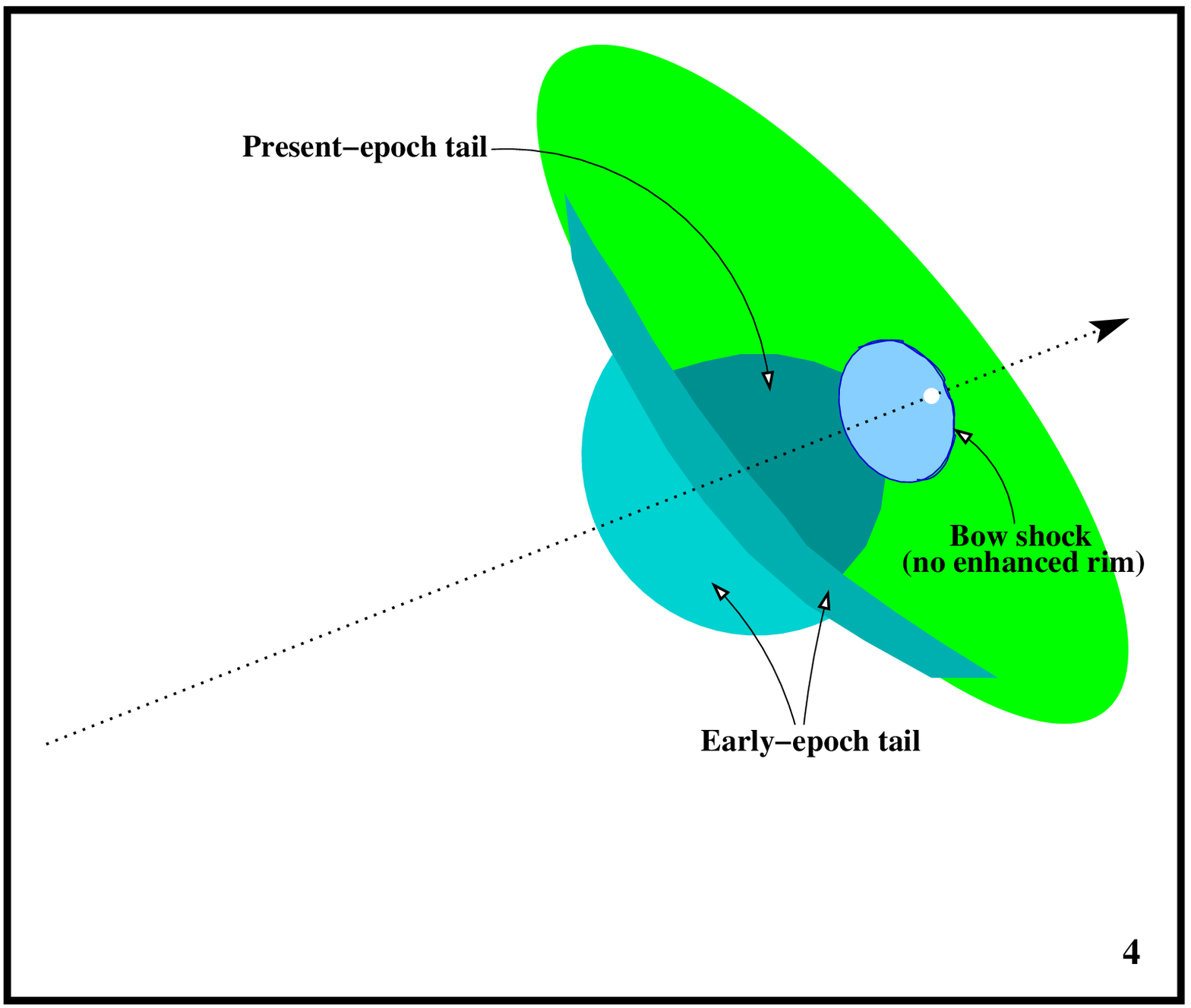}}
\caption{A four-panel {\it cartoon} history of the interaction between
  \sh\@ and the ISM.  The first two panels show GD~561 (white dot) and
  \sh\@ (circular blue shell) approaching the \ion{H}{1} cloud (green
  ellipse).  GD~561 is located at approximately the center of \sh\@.
  The shell is enhanced at the leading edge where a bow shock
  separates the nebula from the ISM.  Material stripped from the
  leading edge is deposited downstream into a growing tail (cyan
  ellipse).  The tail formed during the pre-cloud interaction is
  referred to as the {\it early-epoch} tail.  The third panel shows
  \sh\@ after it has fully entered the cloud.  The increased ram
  pressure at the leading edge has distorted the bow shock, slowed the
  progression of the upstream portion of \sh\@ (relative to GD~561),
  and increased stripping.  The early-epoch tail now has two parts:
  older material which has not yet encountered the cloud, and younger
  material accumulating along the edge of the cloud.  We also see a
  {\it present-epoch} tail, consisting of material stripped since
  \sh\@ entered the cloud. (Increased tail density is indicated by
  darker shades of cyan.) The fourth panel represents the present-day
  picture.  GD~561 is close to the leading edge of a flattened \sh\@.
  The bow shock is now less prominent, as much of the leading-edge
  material has been strewn downstream.}
\label{CartoonHistory}
\end{figure}

The Faraday rotation revealed in our 1420~MHz polarization images is
naturally explained by an interaction between \sh\@ and the ISM: we
see tail-like structures lagging downstream of \sh\@ and a deflection
of the ISM magnetic field at the leading edge of \sh\@.  Indeed, we
are catching the likely interaction during a particularly interesting
epoch in its history.  The projected trajectory of GD~561 indicates
that GD~561 and \sh\@ entered the \ion{H}{1} cloud $\sim$27,000~yr ago
(see Figures~8, 9).  The increase in ambient density and magnetic
field in the cloud, compared to the surrounding ISM, has likely
ramped-up the severity of the interaction during this time.
Nevertheless, to explain all the observed Faraday-rotation structures
in our images, we need to consider the full history of the interaction
between \sh\@ and the ISM. (Refer to Figure~\ref{CartoonHistory} for a
pictorial representation of the following discussion.)

Prior to entering the \ion{H}{1} cloud, GD~561 and \sh\@ were moving
through a relatively low-density environment ($n_H \sim
0.3$~$\rm{cm}^{-3}$ for the mean ISM at $z \approx 130$~pc, e.g.,
\citealt*{DickeyL1990}).  Ram-pressure stripping from the head of the
bow shock during this earlier epoch produced a tail behind \sh\@.  We
still see this {\it early-epoch} tail today.  The oldest material
continues to move toward the \ion{H}{1} cloud, forming what we call
region 2.  The width of this part of the tail (orthogonal to the
projected trajectory) is $\sim$$0.3\arcdeg$, or $\sim$$1.5$ times the
major-axis length of \sh\@.  This is about what we would expect based
on simulations \citep{WareingZO2007main}.  Assuming the tail material
in region 2 moves with a speed lagging that of GD~561 by
$\gtrsim$15~$\kms$ \citep[see][]{Wareing+2007mira}, and taking the
separation between the eastern edge of region 2 and the present
position of GD~561 as the maximum extent of the tail ($3.1 \pm 0.9$~pc
at $\rm{d} = 415 \pm 120$~pc), we estimate a lifetime for the PN-ISM
interaction $\lesssim$$(2.0 \pm 0.3) \times 10^5$~yr (where the
uncertainty arises from the proper-motion uncertainty for GD~561 given
in Table~\ref{thestar}).  A direct measurement of the velocity of the
tail material in region 2 would further constrain this estimate.

Newer material in the early-epoch tail, i.e., material stripped more
recently than that seen in region 2, is responsible for the arms of
the $S$-shaped structure.  As (ionized) tail material approaches the
\ion{H}{1} cloud, it encounters a rapidly increasing magnetic pressure
($P_{B} = B^2 / 8\pi$) along the trajectory of GD~561, and is
deflected into a likely turbulent flow parallel to the cloud's
magnetic field.  The LOS component of the cloud's field broadens the
depth of the tail, while the on-sky component broadens the width.
Since the arms of the $S$-shaped structure align precisely with the
edge of the cloud, we infer that the cloud's on-sky field component
has a position angle very similar to the cloud's major axis
($\sim$$45\arcdeg$, east of north).  The higher mean RM found for the
arms compared to region 2 ($-16 \pm 5$\,$\rm{rad}\,\rm{m}^{-2}$
vs.\ $-6 \pm 2$\,$\rm{rad}\,\rm{m}^{-2}$) is not surprising, since the
arms have a higher LOS field, longer path length, and likely higher
electron density (as tail material accumulates).  Differences in the
appearance of the north and south arms may arise due to a
north-to-south gradient in the cloud's magnetic-field strength
(perhaps matching the observed hydrogen-density gradient).  In
particular, a weaker field in the south could allow material to spread
more easily across field lines, yielding a wider arm with lower mean
density.

When \sh\@ entered the cloud, the ram pressure of the ISM at its
leading edge ($P_{\rm ram} = \rho_{\rm{ISM}} v^{2}_{\rm{S}}$)
increased by a factor $\sim$10 (compare $n_H \sim 0.3$~$\rm{cm}^{-3}$
outside the cloud to $n_H \sim 3$~$\rm{cm}^{-3}$ for the central ridge
of the cloud). (The magnetic pressure is much lower than the ram
pressure, even for a compressed cloud field at the leading edge of
\sh\@.) The result was likely a factor $\sim$3 decrease in the forward
speed of \sh\@, as the nebula found a new pressure balance
(viz.\ $\rho_{w} v^{2}_{w} = \rho_{\rm{ISM}} v^{2}_{\rm{S}}$), and an
increased disruption of the bow shock.  Over $\sim$20,000~yr (allowing
\sh\@ to first fully enter the cloud), a factor $\sim$3 difference
between the speeds of GD~561 and \sh\@ amounts to an angular
separation (at $\rm{d} = 415 \pm 120$~pc) of $0.07\arcdeg \pm
0.02\arcdeg$.  This is consistent with the observed separation at
present between GD~561 and the center of the optical (i.e., combined
R-band and B-band) nebula.  As for the bow shock, the increased
disruption led to increased and more violent stripping.  This in turn
led to a denser and more turbulent tail \citep{WareingZO2007main}.  We
see this {\it present-epoch} tail as the wide bridge of the $S$-shaped
structure.  The large fluctuations in RM throughout most of the bridge
are consistent with tail turbulence in the form of vortices
\citep*[e.g.,][]{WareingZO2007vort}.  Region 4 along the projected
trajectory of GD~561 has relatively low turbulence compared to its
surroundings in the bridge.  This is not unexpected, as some
simulations show a ``quiet'' zone in the tail immediately behind the
interacting PN \citep[e.g.,][]{WareingZO2007main}.  Also, region 4 is
offset slightly north of the center of the bridge; i.e., the
present-epoch tail extends farther south of the projected path of
\sh\@ than it does north.  This is reasonable, since the ambient
cloud, which surrounds and (somewhat) confines the tail, decreases in
density north-to-south.

Can the trajectory of GD~561 account for the deflection of the ambient
magnetic field from into-the-sky east of \sh\@ to out-of-the-sky
immediately west of \sh\@?  The orientation and RMs for the $S$-shaped
structure indicate that the ambient field in the cloud has an on-sky
component oriented approximately northeast-southwest and a LOS
component directed into the sky.  In contrast, the space velocity of
GD~561 (and \sh\@) has an on-sky component directed
south-southeast-to-north-northwest and (smaller) LOS component
directed out-of the sky.  Thus, \sh\@ intersects the ambient field at
nearly a right angle.  Such an intersection should yield a maximum
deflection \citep*[e.g.,][]{SokerD1997}.  Furthermore, since \sh\@ has
a velocity component out of the sky, it seems natural that sightlines
near the leading edge of \sh\@ (region 7 and the westernmost portion
of region 6) reveal a small out-of-the-sky field component.  Full
magnetohydrodynamic simulations are needed in order to better judge
the observed behavior of the ISM field during the PN-ISM interaction.

\subsection{Resolving Discrepancies with the PN Designation \label{ResolvingDiscrepancies}}

Our radio results for \sh\@ favor an interacting PN interpretation
over a Str{\"o}mgren-zone interpretation.  If an interacting PN is
indeed the correct interpretation, then \sh\@ is in the late stages of
an interaction with an \ion{H}{1} cloud.  As a result, some of the
usual observational markers for interacting PNe may be muddled or
missing.  For example, \sh\@ does not have a distinct bow shock.  This
fact should not be considered surprising, as much of the nebular
material originally at the leading edge of the interaction has been
strewn downstream into the tail.  The R-band cleft-hoof structure may
correspond to material still moving downstream toward the tail.  The
relatively narrow widths of the emission lines \citep{Madsen+2006} and
the radial velocity offset between the ionized gas and GD~561
\citep{Frew2008PhDT} are also peculiarities for \sh\@.  However, these
too can be explained by the interaction between \sh\@ and the cloud:
the expansion of the PN shell has been halted at the leading edge and
restricted elsewhere, and the overall motion of \sh\@ has been slowed
in the cloud by a factor $\sim$3 compared to GD~561.

If GD~561 evolved via the standard AGB route, its low mass (see
Table~\ref{thestar}) implies a post-AGB age $>$$10^6$~yr
\citep{Blocker1995b}.  This is a factor $>$5 times larger than the
interaction timescale we estimate for \sh\@.  The age gap is similarly
problematic if we consider a post-early-AGB evolution; i.e., evolution
following the early termination of the AGB stage
(\citealt*{DormanRO1993}; \citealt{Blocker1995b}).  However, if GD~561
evolved via a red-giant-branch (RGB) route, its estimated (post-RGB)
age is $\sim$$10^5$~yr \citep{Driebe+1998}.  No observed PNe have been
unambiguously identified as post-RGB objects
\citep[e.g.,][]{DeMarco2009,Hall+2013}, but RGB remnants, i.e., helium
white dwarfs, are an expected product of close binary evolution and
are estimated to make-up $\sim$4\% of PN central stars
(\citealt*{NieWN2012}; \citealt{DeMarco+2013}).  Though
\citet{Good+2005} find no evidence of binarity in GD~561's radial
velocity, and there is no evidence of an infrared excess for GD~561
\citep[see][]{Zacharias+2013}, the existence of a companion cannot be
excluded.  On account of our radio results for \sh\@, it is perhaps
reasonable to include the post-RGB scenario in PN-ISM interaction
simulations.

\subsection{Origin of the $I$-Emission Plateau \label{OriginOfPlateau}}

Our 1420~MHz $I$ image reveals a low-surface-brightness plateau which
appears to emanate from the $I$-emission halo around \sh\@ and run
$0.30\arcdeg \pm 0.05\arcdeg$ southeast from edge of the \ion{H}{1}
cloud (see Figure~\ref{AllInOne}).  One possible explanation is that
the plateau traces the extent to which ionizing photons from GD~561
penetrate the surrounding ISM, after escaping the eastern boundary of
the \ion{H}{1} cloud.  To evaluate this, we consider the electron
density in the plateau.  The mean brightness temperature over the
plateau is $T_b = 0.045 \pm 0.004$~K.  Assuming $T_e = 10000 \pm
2000$~K, we estimate a mean EM for the plateau $29 \pm
4$~$\rm{cm}^{-6}$\,pc.  Now, using an angle-equivalent path length
$0.5\arcdeg \pm 0.2\arcdeg$ (which translates at $\rm{d} = 415 \pm
120$~pc to a physical path length $3.6 \pm 1.8$~pc), we derive a mean
electron density for the plateau $n_e = 2.8 \pm 0.7$~$\rm{cm}^{-3}$.
This is much larger than the expected $n_H \sim 0.3$~$\rm{cm}^{-3}$
outside the cloud.

Another possibility is that the plateau was formed by an outflow of
ionized material from the \ion{H}{1} cloud.  \sh\@ entered the cloud
$\sim$27,000~yr ago, ionizing (via GD~561) ambient material and
depositing additional ionized material in its wake.  The thermal
pressure in this part of the cloud increased correspondingly ($P_{\rm
  th} = n k T$), and the approximate pressure balance between the
cloud and surrounding ISM, particularly at the boundary behind \sh\@,
was broken.  To cover $0.30\arcdeg \pm 0.05\arcdeg$ ($2.2 \pm 0.7$~pc
at $\rm{d} = 415 \pm 120$~pc) in $\sim$27,000~yr, the resulting
outflow speed must be $80 \pm 30$~$\kms$.  This is $\sim$2 times
higher than that seen elsewhere for outflows produced by \ion{H}{2}
regions breaching the boundary of a confining cloud
\citep[e.g.,][]{Depree+1994}.  We can account for most (or perhaps
all) of the difference by increasing the flow time.  Given the large
extent of the GD~561 Str{\"o}mgren zone in the low-density ISM, it is
quite plausible that the outflow started well before \sh\@ intercepted
the cloud.  A direct measurement of the motion of the ionized material
in the plateau is necessary to fairly evaluate the outflow model.

Is it surprising that we observe no discernible Faraday rotation on
the $I$-emission plateau (outside its superposition with the bridge of
the $S$-shaped structure)?  In short, no.  Thermal emission is
proportional to the square of the electron density, while Faraday
rotation is proportional to the product of electron density and LOS
magnetic field.  The electron density in the plateau ($n_e = 2.8 \pm
0.7$~$\rm{cm}^{-3}$) is not high enough, when combined with the
relatively low LOS field away from the \ion{H}{1} cloud, to rotate the
polarization angle of the diffuse background emission by (an
observable) $\gtrsim$$3\arcdeg$.

\section{CONCLUSIONS \label{concl}}

Here we give a summary of our results and conclusions:

1.  We have presented 1420~MHz polarization, 1420~MHz total intensity
($I$), and neutral hydrogen (\ion{H}{1}) images of the region around
\sh\@.  A superposition of the images shows a physical interconnection
between the ionized and neutral structures.

2.  The \ion{H}{1} images indicate that \sh\@ sits presently at the
center of an \ion{H}{1} cloud.  The cloud is $\sim$$1.2\arcdeg$-long
and $\sim$$0.4\arcdeg$-wide, and oriented approximately
northeast-southwest.  Assuming the cloud is optically thin at 21~cm,
we estimate a peak neutral hydrogen density $n_H = 4 \pm
2$~$\rm{cm}^{-3}$.

3.  The $I$ image reveals thermal emission peaks coincident with the
\sh\@ R-band emission.  We estimate electron densities in the peaks
$n_e = 11 \pm 3$~$\rm{cm}^{-3}$.  The $I$ image also reveals a
low-surface-brightness ``halo'' around \sh\@, and a
low-surface-brightness ``plateau'' running $0.30\arcdeg \pm
0.05\arcdeg$ southeast from the edge of the \ion{H}{1} cloud.

4.  The polarization images reveal a high-contrast Faraday-rotation
structure oriented approximately northeast-southwest and merging at
its midpoint with the downstream edge of \sh\@.  The ``arms'' of this
structure run precisely along the eastern edge of the \ion{H}{1}
cloud.  The polarization images also reveal lower-contrast
Faraday-rotation structures along the projected trajectory of GD~561.

5.  Our RM analysis indicates that the ISM magnetic field around \sh\@
is generally directed into the sky, but deflected at the leading edge
of \sh\@ such that we find a small out-of-the-sky component at
approximately the position of GD~561.  We estimate for the Faraday
rotation immediately behind \sh\@ $\rm{RM} = -44 \pm
5$\,$\rm{rad}\,\rm{m}^{-2}$.  Combining this with an estimate of the
electron density behind \sh\@, we derive a LOS field component $B_{\|}
= 8 \pm 2$~$\mu\rm{G}$.  If this value is representative of the
ambient field in the \ion{H}{1} cloud, and the field lies as much in
the plane of the sky as along the LOS, we estimate a total cloud field
$B = 11 \pm 3$~$\mu\rm{G}$.  This estimate is consistent with the
statistical mean for diffuse \ion{H}{1} sheets and filaments.

6.  Our radio results are inconsistent with a Str{\"o}mgren-zone
interpretation for \sh\@: we see a mean electron density for \sh\@
($n_e = 8 \pm 2$~$\rm{cm}^{-3}$) that is a factor $\sim$2 higher than
the peak hydrogen density in the \ion{H}{1} cloud, prominent
Faraday-rotation structures downstream of \sh\@, and a deflected
magnetic field at the leading edge of \sh\@.  On the other hand, our
radio results are consistent with a PN-ISM interaction.

7.  Our space-velocity analysis for GD~561 indicates that GD~561 and
\sh\@ entered the \ion{H}{1} cloud $\sim$27,000~yr ago.  The increased
severity of the PN-ISM interaction in this time has likely led to a
factor $\sim$3 decrease in the forward speed of \sh\@ and an increased
disruption of the bow shock.

8. The transition from outside to inside the \ion{H}{1} cloud explains
the overall shape of the high-contrast Faraday-rotation structure: The
arms of the structure likely represent an {\it early-epoch} tail;
i.e., material stripped before \sh\@ reached the cloud, and now
spread-out along the cloud's edge.  The broad mid-region of the
structure represents the {\it present-epoch} tail; i.e., material
stripped since \sh\@ entered the cloud, and now sitting immediately
behind \sh\@.

9. We estimate a PN-ISM interaction timescale $\lesssim$$2.0 \times
10^5$~yr, consistent with a post-RGB evolutionary path for the
low-mass GD~561.  We suggest that the post-RGB scenario be added to
simulations of the PN-ISM interaction.

10. Even though \sh\@ itself is not a Str{\"o}mgren zone, the
$I$-emission halo surrounding \sh\@ likely does correspond to the
GD~561 Str{\"o}mgren zone: the extent of the halo is largest where the
\ion{H}{1} cloud is most diffuse, and smallest where the cloud is most
dense.  The $I$-emission plateau may be an outflow of material from
the ionized cloud region behind \sh\@.

\acknowledgements

ACKNOWLEDGMENTS.  We thank the anonymous referees for constructive
reviews of the paper and for comments helpful in the preparation of
the revised manuscript.  RRR thanks the ESL committee at Okanagan
College for approving research leave.  RRR also thanks Travis Rector
for sharing his KPNO 4-m data for \sh\@ and for useful discussions.
The Dominion Radio Astrophysical Observatory is operated as a national
facility by the National Research Council Canada.  This research is
based in part on observations with the 100-m telescope of the MPIfR at
Effelsberg.  The Second Palomar Observatory Sky Survey (POSS-II) was
made by the California Institute of Technology with funds from the
National Science Foundation, the National Geographic Society, the
Sloan Foundation, the Samuel Oschin Foundation, and the Eastman Kodak
Corporation.  This research has made use of the VizieR catalogue
access tool, CDS, Strasbourg, France.


\bibliographystyle{apj}
\bibliography{sh174.bib}


\begin{deluxetable}{l c c c c}
\tabletypesize{\scriptsize}
\tablecaption{Properties of GD~561\label{thestar}}
\tablewidth{0pt}
\tablehead{
  \colhead{Parameter} &
  \colhead{} &
  \colhead{Value} &
  \colhead{} &
  \colhead{Reference}
}
\startdata
Equatorial Coordinates (J2000)              &  & 23 45 02.254, +80 56 59.72                                                                 &  & 1 \\
Galactic Coordinates (deg)                  &  & 120.2169, +18.4302                                                                         &  & 1 \\
Proper Motion ($\masyr$)                    &  & $23.7 \pm 2.2$\tablenotemark{a}                                                            &  & 1 \\
Radial Velocity ($\kms$)                    &  & $-12.5 \pm 0.7$\tablenotemark{b}                                                           &  & 2 \\
                                            &  &                                                                                            &  & \\
Spectral Classification                     &  & DAO                                                                                        &  & 3 \\
$T_{\rm{eff}}$ (K)                            &  & $64354 \pm 2909$ (Ba), $75627 \pm 4953$ (Ly), $74160$ (Ba), $69000$ (Ly+)\tablenotemark{c} &  & 4,4,5,6 \\
log\,$g$ ($\cmss$)                          &  & $6.94 \pm 0.16$ (Ba), $6.64 \pm 0.06$ (Ly), $7.16$ (Ba), $6.70$ (Ly+)\tablenotemark{c}     &  & 4,4,5,6 \\
Mass ($\Msol$)                              &  & $0.464 \pm 0.029$ (Ba), $0.442 \pm 0.023$ (Ly), $0.48$ (Ba), $0.40$ (Ly+)\tablenotemark{c} &  & 2,2,5,6 \\
                                            &  &                                                                                            &  & \\
Visual Magnitude ($V$)                      &  & $14.59$                                                                                    &  & 1 \\
Interstellar Extinction ($E(B-V)$)          &  & $0.089$                                                                                    &  & 4 \\
%
                                            &  &                                                                                            &  & \\
Distance (pc)                               &  & $415 \pm 120$\tablenotemark{d}                                                             &  & \\
\enddata
\tablecomments{Units of right ascension are hours, minutes, and
seconds, and units of declination are degrees, arcminutes, and
arcseconds; mas $\equiv$ milliarcseconds.}
\tablenotetext{a}{Equatorial components for the proper motion:
  $\mu_{\alpha} = -23.5 \pm 2.1$, $\mu_{\delta} = +2.7 \pm 2.5$
  $\masyr$.}
\tablenotetext{b}{The radial velocity includes the gravitational
  redshift, which we estimate, using the expression given in
  \citet{Barstow+2005} and the range of tabulated values for
  $T_{\rm{eff}}$ and mass, to be $5.1 \pm 0.8$~$\kms$.  The
  gravitational-redshift-corrected radial velocity is then $-17.6 \pm
  1.1$~$\kms$, where the uncertainty is the rss of the estimated
  errors in the (uncorrected) radial velocity and gravitational
  redshift.  We use this corrected value in our derivation of the
  space velocity of GD~561.}
\tablenotetext{c}{Optical and far ultraviolet (FUV) spectroscopic
  analyses give systematically different values for $T_{\rm{eff}}$ and
  log\,$g$.  We use Ba to denote an optical analysis using the Balmer
  lines, Ly to denote an FUV analysis using only the Lyman lines, and
  Ly+ to denote an FUV analysis using several lines including the
  Lyman lines.  The mass estimates, which are determined by comparing
  the estimated $T_{\rm{eff}}$ and log\,$g$ values with evolutionary
  models, also show a systematic variation.}
\tablenotetext{d}{``Gravity'' distance \citep[see,
    e.g.,][]{Napiwotzki2001} estimated using the tabulated values for
  $T_{\rm{eff}}$, log\,$g$, and mass, as well as those for $V$ and
  $E(B-V)$.  The large error range reflects largely the systematic
  differences for the values of $T_{\rm{eff}}$, log\,$g$, and mass:
  Lyman-line-estimated values give distances at the high end of the
  range, while Balmer-line-estimated values give distances at the low
  end.}
\tablerefs{
  1. UCAC4 (\citealt{Zacharias+2013});\phn
  2. \citealt{Good+2005};\phn
  3. \citealt{Bergeron+1994};\phn
  4. \citealt{Good+2004};\phn
  5. \citealt{Gianninas+2010};\phn
  6. \citealt{Ziegler+2012}.
  }
\end{deluxetable}

\begin{deluxetable}{l c c c c c c c}
\tabletypesize{\scriptsize}
\tablecaption{Faraday-Rotation Model Summary\label{RMTable}}
\tablewidth{0pt}
\tablehead{
  \colhead{Region} &
  \colhead{} &
  \colhead{BPI (K)} &
  \colhead{BPA} &
  \colhead{FPI (K)} &
  \colhead{FPA} &
  \colhead{RM ($\rm{rad}\,\rm{m}^{-2}$)} &
  \colhead{Beam Depol.}
}
\startdata
2    &  &  $0.14 \pm 0.02$ & $-30\arcdeg \pm 10\arcdeg$ & $0.13 \pm 0.02$ & $-15\arcdeg \pm 10\arcdeg$ &  \phn$-6 \pm 2$ & $30\% \pm 5\%$ \\
4    &  &  $0.14 \pm 0.02$ & $-30\arcdeg \pm 10\arcdeg$ & $0.13 \pm 0.02$ & $-15\arcdeg \pm 10\arcdeg$ &  $-44 \pm 5$    & $25\% \pm 5\%$ \\
6    &  &  $0.14 \pm 0.02$ & $-30\arcdeg \pm 10\arcdeg$ & $0.13 \pm 0.02$ & $-15\arcdeg \pm 10\arcdeg$ &  \phn$-1 \pm 2$ & $15\% \pm 5\%$ \\
7    &  &  $0.14 \pm 0.02$ & $-30\arcdeg \pm 10\arcdeg$ & $0.13 \pm 0.02$ & $-15\arcdeg \pm 10\arcdeg$ &  \phn$+8 \pm 2$ & \phn$5\% \pm 5\%$ \\
Arms &  &  $0.14 \pm 0.02$ & $-30\arcdeg \pm 10\arcdeg$ & $0.13 \pm 0.02$ & $-15\arcdeg \pm 10\arcdeg$ &  $-16 \pm 5$    & $50\% \pm 8\%$ \\
\enddata
\tablecomments{BPI, FPI $\equiv$ background and foreground polarized
  intensity, respectively; BPA, FPA $\equiv$ background and foreground
  polarization angle, respectively.}
\end{deluxetable}


\end{document}